\documentclass[longnamesfirst]{emulateapj}
\usepackage{natbib}
\usepackage{apjfonts}
\usepackage{xspace}
\tighten

\def\pnorm{~photons\,keV$^{-1}$\,cm$^{-2}$\,s$^{-1}$ at 1\,keV}

\def\obj{\hbox{NGC\,1068}\xspace}
\def\nustar{{\it NuSTAR}\xspace}
\def\chandra{{\it Chandra}\xspace}
\def\xmm{{\it XMM-Newton}\xspace}
\def\swift{{\it Swift}\xspace}
\def\suzaku{{\it Suzaku}\xspace}
\def\bepposax{{\it BeppoSAX}\xspace}

\def\xspec{{\sc xspec}\xspace}

\slugcomment{The Astrophysical Journal (submitted, 11/03/14)}
\shortauthors{BAUER ET AL.}
\shorttitle{Multiple Reflections in NGC\,1068}

\begin{document}

\title{NuSTAR Spectroscopy of Multi-Component X-ray Reflection from NGC\,1068}

\author{
Franz~E.~Bauer,\altaffilmark{1,2,3,4} 
Patricia Ar\'{e}valo,\altaffilmark{5,3} 
Dominic J. Walton,\altaffilmark{6} 
Michael J. Koss,\altaffilmark{7,8} 
Simonetta Puccetti,\altaffilmark{9,10} 
Poshak Gandhi,\altaffilmark{11} 
Daniel Stern,\altaffilmark{6} 
David M. Alexander,\altaffilmark{12} 
Mislav Balokovi\'{c},\altaffilmark{13} 
Steve E. Boggs,\altaffilmark{14}
William N. Brandt,\altaffilmark{15,16} 
Murray Brightman,\altaffilmark{13} 
Finn E. Christensen,\altaffilmark{17}
Andrea Comastri,\altaffilmark{18} 
William W. Craig,\altaffilmark{14,19}
Agnese Del Moro,\altaffilmark{12} 
Charles J. Hailey,\altaffilmark{20}
Fiona A. Harrison,\altaffilmark{13}
Ryan Hickox,\altaffilmark{21}
Bin Luo,\altaffilmark{15} 
Craig B. Markwardt,\altaffilmark{22} 
Andrea Marinucci,\altaffilmark{23}
Giorgio Matt,\altaffilmark{23}
Jane R. Rigby,\altaffilmark{22} 
Elizabeth Rivers,\altaffilmark{13} 
Cristian Saez,\altaffilmark{24} 
Ezequiel Treister,\altaffilmark{25,3} 
C. Megan Urry,\altaffilmark{26} 
and William W. Zhang.\altaffilmark{22} 
}

\altaffiltext{1}{Pontificia Universidad Cat\'{o}lica de Chile, Instituto de Astrof\'{\i}sica, Casilla 306, Santiago 22, Chile} 
\altaffiltext{2}{Millenium Institute of Astrophysics, Santiago, Chile} 
\altaffiltext{3}{EMBIGGEN Anillo, Concepci\'{o}n, Chile} 
\altaffiltext{4}{Space Science Institute, 4750 Walnut Street, Suite 205, Boulder, Colorado 80301}
\altaffiltext{5}{Instituto de F\'{\i}sica y Astronom\'{a}, Facultad de Ciencias, Universidad de Valpara\'{\i}so, Gran Bretana N 1111, Playa Ancha, Valpara\'{\i}so, Chile} 
\altaffiltext{6}{Jet Propulsion Laboratory, California Institute of Technology, 4800 Oak Grove Drive, Pasadena, CA 91109, USA} 
\altaffiltext{7}{Institute for Astronomy, Department of Physics, ETH Zurich, Wolfgang-Pauli-Strasse 27, CH-8093 Zurich, Switzerland} 
\altaffiltext{8}{SNSF Ambizione Postdoctoral Fellow} 
\altaffiltext{9}{ASDC-ASI, Via del Politecnico, I-00133 Roma, Italy} 
\altaffiltext{10}{INAF-Osservatorio Astronomico di Roma, via Frascati 33, I-00040 Monte Porzio Catone (RM), Italy} 
\altaffiltext{11}{School of Physics and Astronomy, University of Southampton, Highfield, Southampton SO17 1BJ, UK} 
\altaffiltext{12}{Department of Physics, Durham University, South Road, Durham, DH1 3LE, UK} 
\altaffiltext{13}{Cahill Center for Astronomy and Astrophysics, California Institute of Technology, Pasadena, CA 91125, USA} 
\altaffiltext{14}{Space Sciences Laboratory, University of California, Berkeley, CA 94720, USA} 
\altaffiltext{15}{Department of Astronomy and Astrophysics, The Pennsylvania State University, 525 Davey Lab, University Park, PA 16802, USA} 
\altaffiltext{16}{Institute for Gravitation and the Cosmos, The Pennsylvania State University, University Park, PA 16802, USA} 
\altaffiltext{17}{DTU Space, National Space Institute, Technical University of Denmark, Elektrovej 327, 2800 Lyngby, Denmark} 
\altaffiltext{18}{INAF-Osservatorio Astronomico di Bologna, via Ranzani 1, I-40127 Bologna, Italy} 
\altaffiltext{19}{Lawrence Livermore National Laboratory, Livermore, CA 945503, USA} 
\altaffiltext{20}{Columbia Astrophysics Laboratory, Columbia University, New York, NY 10027, USA} 
\altaffiltext{21}{Department of Physics and Astronomy, Dartmouth College, 6127 Wilder Laboratory, Hanover, NH 03755, USA} 
\altaffiltext{22}{NASA Goddard Space Flight Center, Greenbelt, MD 20771, USA} 
\altaffiltext{23}{Dipartimento di Matematica e Fisica, Universit\'{a} degli Studi Roma Tre, via della Vasca Navale 84, I-00146 Roma, Italy} 
\altaffiltext{24}{Department of Astronomy, University of Maryland, College Park, MD 20742, USA} 
\altaffiltext{25}{Departamento de Astronom\'{\i}a Universidad de Concepci\'{o}n, Casilla 160-C, Concepci\'{o}n, Chile} 
\altaffiltext{26}{Department of Physics and Yale Center for Astronomy and Astrophysics, Yale University, New Haven, CT 06520-8120, USA} 

\begin{abstract}
We report on observations of NGC\,1068 with \nustar{}, which provide
the best constraints to date on its $>10$~keV spectral shape. The
\nustar{} data are consistent with past instruments, with no strong
continuum or line variability over the past two decades, consistent
with its classification as a Compton-thick AGN. The combined
\nustar{}, \chandra{}, \xmm{}, and \swift{} BAT spectral dataset
offers new insights into the complex secondary emission seen instead
of the completely obscured transmitted nuclear continuum. The critical
combination of the high signal-to-noise \nustar{} data and the
decomposition of the nuclear and extranuclear emission with \chandra{}
allow us to break several model degeneracies and greatly aid physical
interpretation. When modeled as a monolithic (i.e., a single $N_{\rm
  H}$) reflector, none of the common Compton-reflection models are
able to match the neutral fluorescence lines and broad spectral shape
of the Compton reflection without requiring unrealistic physical
parameters (e.g., large Fe overabundances, inconsistent viewing
angles, poor fits to the spatially resolved spectra). A
multi-component reflector with three distinct column densities (e.g.,
with best-fit values of $N_{\rm H}=1.5\times10^{23}$,
$5\times10^{24}$, and $10^{25}$\,cm$^{-2}$) provides a more reasonable
fit to the spectral lines and Compton hump, with near-solar Fe
abundances. In this model, the higher $N_{\rm H}$ component provides
the bulk of the flux to the Compton hump while the lower $N_{\rm H}$
component produces much of the line emission, effectively decoupling
two key features of Compton reflection. We also find that
$\approx$30\% of the neutral Fe K$\alpha$ line flux arises from
$>$2\arcsec{} ($\approx$140~pc) and is clearly extended, implying that
a significant fraction of the $<$10\,keV reflected component arises
from regions well outside of a parsec-scale torus. These results
likely have ramifications for the interpretation of Compton-thick
spectra from observations with poorer signal-to-noise and/or more
distant objects.
\end{abstract}


\section{Introduction}\label{sec:intro}

At a distance of $\approx$14.4\,Mpc \citep{Tully1988}, NGC\,1068 is
one of the nearest and best-studied Active Galactic Nuclei (AGN). It
is traditionally classified as a Seyfert 2 galaxy, and was the first
type 2 AGN observed to possess polarized optical broad-line emission; these
broad line regions seen only in scattered light are presumably
obscured by a dusty edge-on structure \citep[a.k.a. the
  ``torus'';][]{Antonucci1985, Miller1991}, thereby establishing the
standard orientation-based model of AGN unification as we know it
today \citep{Antonucci1993, Urry1995}. \obj has continued to be an
exceptionally rich source for studying AGN in general and
Compton-thick AGN in particular,\footnote{With a line-of-sight column
  density exceeding $N_{\rm H} = 1.5\times10^{24}$~cm$^{-2}$ and
  therefore optically thick to Compton scattering.} as there are spatially
resolved studies of the AGN structure down to $\approx$0.5--70\,pc over
many critical portions of the electromagnetic spectrum (1\arcsec{} $=$
70 pc at the distance of \obj). In many ways, \obj is considered an
archetype of an obscured AGN.

In terms of its basic properties and structure, H$_{\rm 2}$O megamaser
emission coincident with the nucleus and associated with a thin disk,
has constrained the supermassive black hole (SMBH) mass at the center
of \obj to be \hbox{$\approx1\times10^{7}$ M$_{\odot}$} within 0.65
pc, although the observed deviations from Keplerian rotation leave
some ambiguity about the overall mass distribution
\citep[e.g.,][]{Greenhill1996, Gallimore2004, Lodoto2003}. A dynamical
virial mass estimate based on the width of the H$_{\beta}$ line,
$\sigma_{\rm H_{\beta}}$, from the scattered ``polarized broad lines''
in the hidden broad line region (BLR) has found a consistent mass of
$(9.0\pm6.6)\times10^{6}$ M$_{\odot}$ \citep[e.g.,][]{Kuo2011}. \obj's
bolometric luminosity has been estimated to be \hbox{$L_{\rm bol}=$
  (6--10)$\times10^{44}$ erg~s$^{-1}$} \citep{Woo2002,
  AlonsoHerrero2011} based on mid-infrared (MIR) spectral modeling
assuming reprocessed AGN emission. Combined with the SMBH mass
estimate, this luminosity approaches $\approx$50--80\% of the
Eddington luminosity, indicating rapid accretion.

Very Long Baseline Interferometry (VLBI) observations of the maser
disk constrain it to lie between radii of 0.6--1.1~pc at a position angle
($PA$) of $\approx -45$\degr{} \citep[east of north;
  e.g.,][]{Greenhill1996}. At centimeter wavelengths, a weak kpc-scale,
steep-spectrum radio jet is seen to extend out from the nucleus,
initially at $PA=12$\degr{} before bending to $PA=30$\degr{} at large
scales \citep[e.g.,][]{Wilson1987, Gallimore1996}. Fainter radio
structures close to the nucleus are also observed to trace both the
maser disk and an inner X-ray-irradiated molecular disk extending out
to $\approx$0.4~pc with a $PA\approx -60$\degr{}
\citep[e.g.,][]{Gallimore2004}.

At MIR wavelengths, a complex obscuring structure has been spatially
resolved in \obj via Keck and VLT interferometry
\citep[e.g.,][]{Bock2000, Jaffe2004} and appears to be comprised of at
least two distinct components \citep{Raban2009, Schartmann2010}. The
first is a $\sim$800\,K, geometrically thin, disk-like structure
extending $\approx$1.35\,pc by 0.45\,pc in size (full-width half
maximum, FWHM) and aligned at $PA=-42$\degr{}, which is likely
associated with the maser disk. The second is a $\sim$300~K, more
flocculent, filamentary, torus-like distribution $\approx$3--4\,pc in
size (FWHM) which has been identified with the traditional torus. The
parameters of the spectral modeling to the overall MIR light are
consistent, with a torus radius of $\approx$2~pc and angular width of
$26^{+6}_{-4}$ deg, a viewing angle of $88^{+2}_{-3}$ deg with respect
to the line-of-sight, and a covering factor of $\approx$25--40\%
\citep{AlonsoHerrero2011}. While no dust reverberation studies have
been published on \obj, the sizes from interferometry are consistent
with the inner radii determined from dust reverberation studies of
type 1 AGN \citep{Suganuma2006, Koshida2014}.

NGC\,1068 also displays a striking extended narrow-line region (NLR)
that is roughly co-spatial with the radio jet and lobe emission
\citep[e.g.,][]{Wilson1987}. The NLR has been extensively
characterized by narrow-band imaging and IFU studies \citep{Evans1991,
  Macchetto1994, Capetti1997, Veilleux2003}. The biconical ionization
cone has been observed out to radii of $\ga$150\arcsec{}, with an
apparent opening angle of $\approx$60\degr{} centered at
PA$\approx$35\degr{}--45\degr{} \citep{Unger1992, Veilleux2003}. The
narrow-line emitting clouds are part of a large-scale, radiatively
accelerated outflow with velocities up to $\approx$3200~km~s$^{-1}$
\citep[e.g.,][]{Cecil1990, Crenshaw2000, Cecil2002}. The morphology of
the NLR seems to primarily trace the edges of the radio lobe,
suggesting that the radio outflow has swept up and compressed the
interstellar gas, giving rise to enhanced line emitting regions. The
energetics of the line emission indicate that it is probably
photoionization dominated \citep{Dopita2002, Groves2004}.  Various
studies have reported strongly non-solar abundances in the ionized gas
of \obj, which either require large over- or underabundances of some
elements \citep[e.g., due to shocks, supernovae pollution of Nitrogen,
  Phosphorus, etc, or that elements like C and Fe are predominantly
  locked in dust grains;][]{Kraemer1998, Oliva2001, Martins2010}, or
can also be explained by multi-component photoionization models with
varying densities \citep[e.g.,][]{Kraemer2000}.

As we now know, the primary AGN continuum of \obj from the optical to
X-rays is completely obscured along our line of sight due to the
relative orientations of the disk and obscuring torus, which has a
column density $N_{\rm H}>10^{25}$~cm$^{-2}$
\citep[e.g.,][]{Matt2000}. Thus the only X-ray emission that we see is
scattered into our line of sight. Past observations have suggested
that there are two ``reflectors'' which contribute to the X-ray
spectrum \citep[e.g.,][]{Matt1997, Guainazzi1999}. The dominant
component is from Compton scattering off the inner ``wall'' of the
neutral obscuring torus, which gives rise to the so-called ``cold''
Compton reflection continuum \citep[e.g.,][]{Lightman1988}. This
emission is characterized by a hard X-ray spectral slope with a peak
around 30\,keV as well as high equivalent-width fluorescent emission
lines \citep[e.g., the dominant 6.4\,keV iron line;][]{Iwasawa1997}. A
second reflector arises from Compton scattering off highly ionized
material associated with the ionization cone. The spectral shape of
the ``warm'' reflector should crudely mirror the intrinsic continuum,
apart from a high-energy cutoff due to Compton downscattering and
potentially significant absorption edges/lines in the spectrum up to a
few keV due to various elements and near $\sim$7\,keV due to Fe
\citep[e.g.,][]{Krolik1995}. Radiative recombination continuum and
line emission (hereafter RRC and RL, respectively) from a broad range
of ions and elements can also be observed in relation to the warm
reflector due to photoionization followed by recombination, radiative
excitation by absorption of continuum radiation and inner shell
fluorescence \citep[][hereafter K14]{Guainazzi1999, Brinkmann2002,
  Kinkhabwala2002, Ogle2003, Kallman2014}. The ionized lines imply
observed outflow velocities of 400--500~km~s$^{-1}$. Photo-ionized
X-ray emission is seen to extend out along the same direction and
opening angle as the radio jet/lobe and NLR \citep{Young2001}.

Past observations of \obj above 10~keV have been limited by available
instrumentation, where statistics were dominated by background.  Here
we report on new observations of \obj between 3--79~keV from
\nustar{}, whose focusing optics reduce background contamination to
unprecedented levels and thus enable a factor of $\ga$10 statistical
improvement over past observations. The \nustar{} data allow the best
characterization of the $>10$~keV spectral shape to date and therefore
stand to yield new insights into the nature of Compton-thick
obscuration.

This paper is organized as follows:
data and reduction methods are briefly detailed in $\S$\ref{sec:data}; 
X-ray spectroscopic constraints for \obj are investigated
in $\S$\ref{sec:spectra}, with particular attention to modeling the
nucleus and galaxy host contamination;
in $\S$\ref{sec:discuss} we discuss some implications of the best fit model;
and finally we summarize and explore future prospects in $\S$\ref{sec:conclude}.
We adopt a Galactic neutral column density of $N_{\rm
  H}=3.0\times10^{20}$~cm$^{-2}$ \citep{Kalberla2005} toward the
direction of \obj and a redshift of 0.00379 \citep{Huchra1999}. Unless
stated otherwise, errors on spectral parameters are for 90\%
confidence, assuming one parameter of interest.

\section{Observational Data and Reduction Methods}\label{sec:data}

Due to the natural limitations of various X-ray instruments in terms
of energy coverage and spectral and angular resolution, our strategy
was to analyze together several high-quality X-ray observations of
\obj obtained by the \nustar{}, \chandra{}, and \xmm{} observatories,
collected between 2000--2013. While \nustar{} and \xmm{} have superior
collecting areas and energy coverage, neither is able to spatially
separate the spectra of the AGN from various sources of host
contamination or resolve some line complexes. We therefore use the
\chandra{} data for these tasks, allowing us to construct the most
robust model to date for the nuclear X-ray spectrum of \obj. We additionally
use \swift{}, \suzaku{}, and \bepposax{} for points of comparison. The
basic parameters of these observations are listed in
Table~\ref{tab:data_xray}. All data were downloaded through the High
Energy Astrophysics Science Archive Research Center {\sc browse}
facility.

For observatories which operate multiple detectors simultaneously
(e.g. EPIC $pn$ and MOS aboard \xmm{}), we model the data from the
different detectors with all parameters tied between the spectra,
incorporating a multiplicative cross-normalization constant in an
attempt to account for any residual internal cross-calibration
uncertainties between the instruments. Likewise, to account for
external cross-calibration discrepancies between observatories, we
also adopt multiplicative cross-normalization constants. The internal
cross-calibration differences for instruments in the same energy range
are generally known to be within $\sim$5\% of unity for all such
missions, while cross-calibration differences both for instruments
with widely different energy ranges and between instruments from
different observatories can be as high as $\sim$30\% (see
Table~\ref{tab:data_xray}). All of the final spectra have been binned
to contain a minimum of 25 counts per bin, sufficient for $\chi^{2}$
minimization.

\begin{deluxetable*}{lrrrrrrr}
\tabletypesize{\scriptsize}
\tablewidth{0pt}
\tablecaption{X-ray Observations\label{tab:data_xray}}
\tablehead{
\colhead{Instrument} & 
\colhead{Date} & 
\colhead{Obsid} &
\colhead{Exp.} & 
\colhead{Energy Band} &
\colhead{Aperture} &
\colhead{Count Rate} &
\colhead{Norm Offset}  
}
\tableheadfrac{0.05}
\startdata
\bepposax MECS  & 1996-12-30 & 5004700100  &     100.8 &   3--10 &   180 &  0.06 & 1.13$\pm$0.02 \\ 
\bepposax PDS   & 1996-12-30 & 5004700100  &     116.6 & 15--140 &  \nodata&  0.27 & 0.70$\pm$0.10\rlap{*} \\ 
\bepposax MECS  & 1998-01-11 & 5004700120  &      37.3 &   3--10 &   180 &  0.04 & 1.11$\pm$0.02 \\ 
\bepposax PDS   & 1998-01-11 & 5004700120  &      31.5 & 15--140 &  \nodata&  0.28 & 0.70$\pm$0.10\rlap{*} \\ 
\chandra{} ACIS-S & 2000-02-21 & 344         &      47.7 & 0.4--8  & 2--75 &  2.09 & 1.04$\pm$0.04 \\ %
\xmm{} pn         & 2000-07-29 & 0111200101  &      32.8 & 0.2--10 &    75 & 12.36 & 1.00 \\ %
\xmm{} pn         & 2000-07-30 & 0111200201  &      28.7 & 0.2--10 &    75 & 12.36 & 1.00 \\ %
\chandra{} HETG HEG/MEG & 2000-12-04 & 332   &      25.7 & 0.3--8  &     2 & 0.031/0.085 & 1.09$\pm$0.06,1.05$\pm$0.06 \\ %
\suzaku XIS     & 2007-02-10 & 701039010   &      61.5 & 0.3--9  &   260 & 0.73 & 1.17$\pm$0.02 \\ %
\suzaku HXD PIN & 2007-02-10 & 701039010   &      38.8 & 15--70  &  \nodata& 0.40 & 1.20$\pm$0.05\rlap{*} \\ %
\chandra{} HETG HEG/MEG & 2008-11-18 & 10816 &      16.2 & 0.8--10/0.4--8 &     2 & 0.029/0.078 & 1.09$\pm$0.06,1.05$\pm$0.06 \\ %
\chandra{} HETG HEG/MEG & 2008-11-19 & 9149  &      89.4 & 0.8--10/0.4--8 &     2 & 0.027/0.077 & 1.09$\pm$0.06,1.05$\pm$0.06 \\ %
\chandra{} HETG HEG/MEG & 2008-11-20 & 10815 &      19.1 & 0.8--10/0.4--8 &     2 & 0.028/0.076 & 1.09$\pm$0.06,1.05$\pm$0.06 \\ %
\chandra{} HETG HEG/MEG & 2008-11-22 & 10817 &      33.2 & 0.8--10/0.4--8 &     2 & 0.028/0.079 & 1.09$\pm$0.06,1.05$\pm$0.06 \\ %
\chandra{} HETG HEG/MEG & 2008-11-25 & 10823 &      34.5 & 0.8--10/0.4--8 &     2 & 0.029/0.077 & 1.09$\pm$0.06,1.05$\pm$0.06 \\ %
\chandra{} HETG HEG/MEG & 2008-11-27 & 9150  &      41.1 & 0.8--10/0.4--8 &     2 & 0.028/0.077 & 1.09$\pm$0.06,1.05$\pm$0.06 \\ %
\chandra{} HETG HEG/MEG & 2008-11-30 & 10829 &      39.6 & 0.8--10/0.4--8 &     2 & 0.027/0.079 & 1.09$\pm$0.06,1.05$\pm$0.06 \\ %
\chandra{} HETG HEG/MEG & 2008-12-03 & 10830 &      44.0 & 0.8--10/0.4--8 &     2 & 0.029/0.078 & 1.09$\pm$0.06,1.05$\pm$0.06 \\ %
\chandra{} HETG HEG/MEG & 2008-12-05 & 9148  &      80.2 & 0.8--10/0.4--8 &     2 & 0.029/0.078 & 1.09$\pm$0.06,1.05$\pm$0.06 \\ %
{\it Swift} BAT (70-month) & 2004--2010 & \nodata & 9250.0 & 14--195 & \nodata & $4.97\times10^{-5}$ & 0.75$\pm$0.10\rlap{*} \\ 
\nustar{} FPMA/FPMB & 2012-12-18 & 60002030002  & 56.9/56.8 &   3--79 &    75 & 0.22/0.21 & 1.11$\pm$0.01 \\ 
{\it Swift} XRT   & 2012-12-19 & 00080252001  &       2.0 & 0.5--10 &    75 & 0.44      & 1.13$\pm$0.25 \\ 
\nustar{} FPMA/FPMB & 2012-12-20 & 60002030004  & 47.8/47.5 &   3--79 &    75 & 0.22/0.21 & 1.11$\pm$0.01 \\ %
\nustar{} FPMA/FPMB & 2012-12-21 & 60002030006  & 19.2/19.4 &   3--79 &    75 & 0.22/0.21 & 1.11$\pm$0.01 \\ %
\enddata
\tablecomments{
{\it Column 1:} Satellite and instrument. 
{\it Column 2:} Starting date of observation.
{\it Column 3:} Observation identification (obsid) number. \chandra{} frametimes were
$\approx$1.9-2.1\,s, except for obsids 355, 356, and 2454 with 3.2\,s and
obsids 365, 9140, and 10937 with 0.3--0.4\,s.
{\it Column 4:} Exposure time in ksec. 
{\it Column 5:} Energy band in keV.
{\it Column 6:} Extraction aperture radius in arcseconds. If a range
is given, then an annular region was extracted.  For the HETG, the
values given represent the half-width of the extraction region in the
cross-dispersion direction.
{\it Column 7:} Count rate in counts s$^{-1}$.
{\it Column 8:} Relative normalization offset with respect to the
pn in the 3--7\,keV band. However, entries denoted by *'s are in fact
relative to combined \nustar{} FPMA/FPMB spectrum in the 20--60\,keV
band.
}
\end{deluxetable*}

\subsection{{\it NuSTAR}}\label{sec:data_nustar}

The \nustar{} observatory is the first focusing satellite with
sensitivity over the broad X-ray energy band from 3--79~keV
\citep{Harrison2013}. It consists of two co-aligned X-ray
optics/detector pairs, with corresponding focal plane modules FPMA and
FPMB, which offer a 12\farcm5$\times$12\farcm5 field-of-view, angular
resolutions of 18\arcsec{} Full Width Half Max (FWHM) and 1\arcmin{}
Half Power Diameter (HPD) over the 3--79~keV X-ray band, and a
characteristic spectral resolution of 400 eV (FWHM) at 10~keV. \obj
was observed by \nustar{} between 2012 December 18--21.

The \nustar{} data were processed using the standard pipeline
\citep[{\sc nupipeline};][]{Perri2013} from the \nustar{} Data
Analysis Software (v1.3.0) within the HEASoft package (v6.15), in
combination with CALDB v20131007. The unfiltered event lists were
screened to reduce internal background at high energies via standard
depth corrections, as well as to remove South Atlantic Anomaly (SAA)
passages. The {\sc nuproducts} program was used to extract data
products from the cleaned event lists for both focal plane modules
FPMA and FPMB.

\begin{figure}[]
\vspace{-1.1in}
\vglue-0.0cm{\hglue0.5cm{\includegraphics[angle=0, width=8cm]{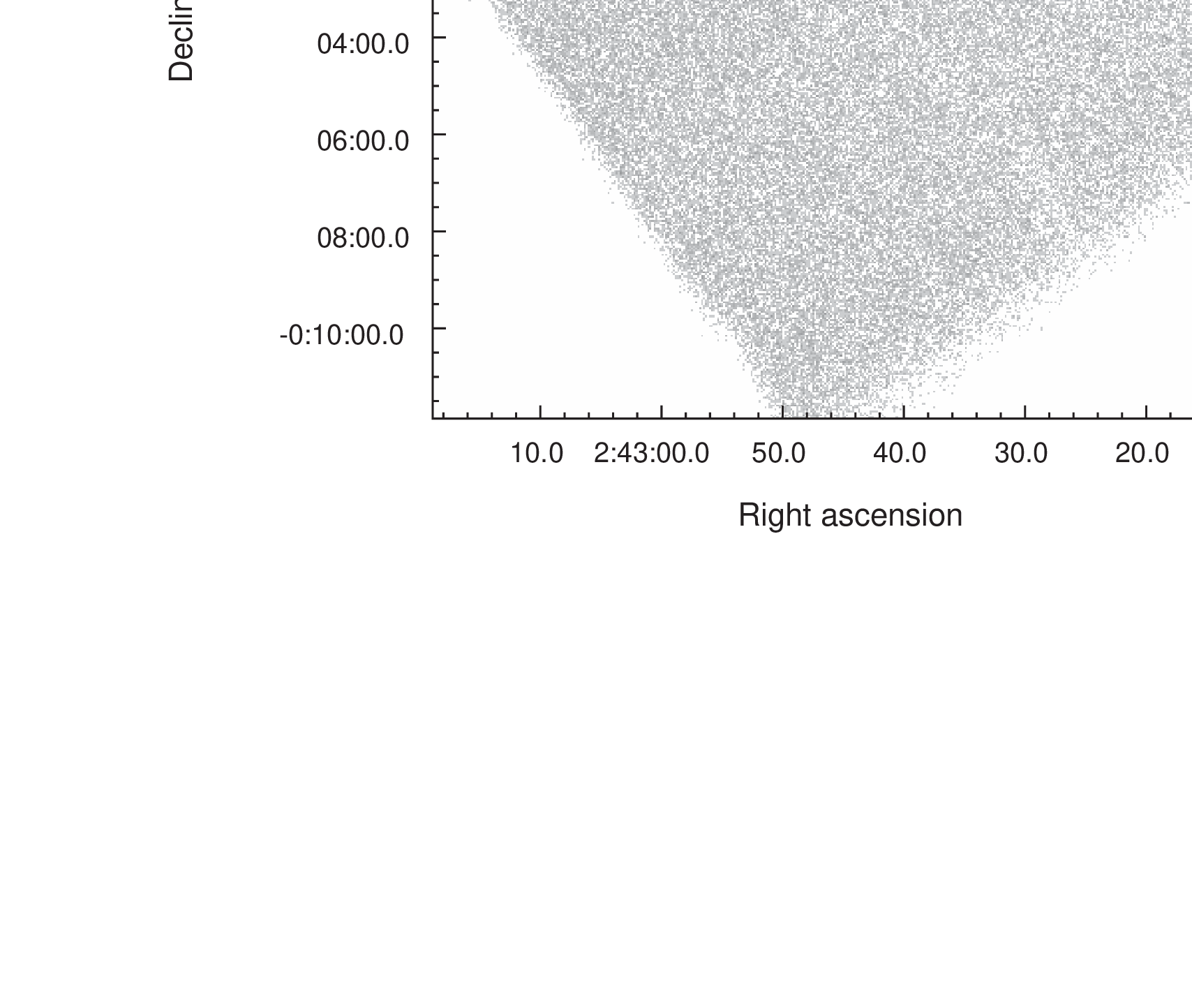}}}
\vglue-0.1cm{\hglue0.5cm{\includegraphics[angle=0, width=8cm]{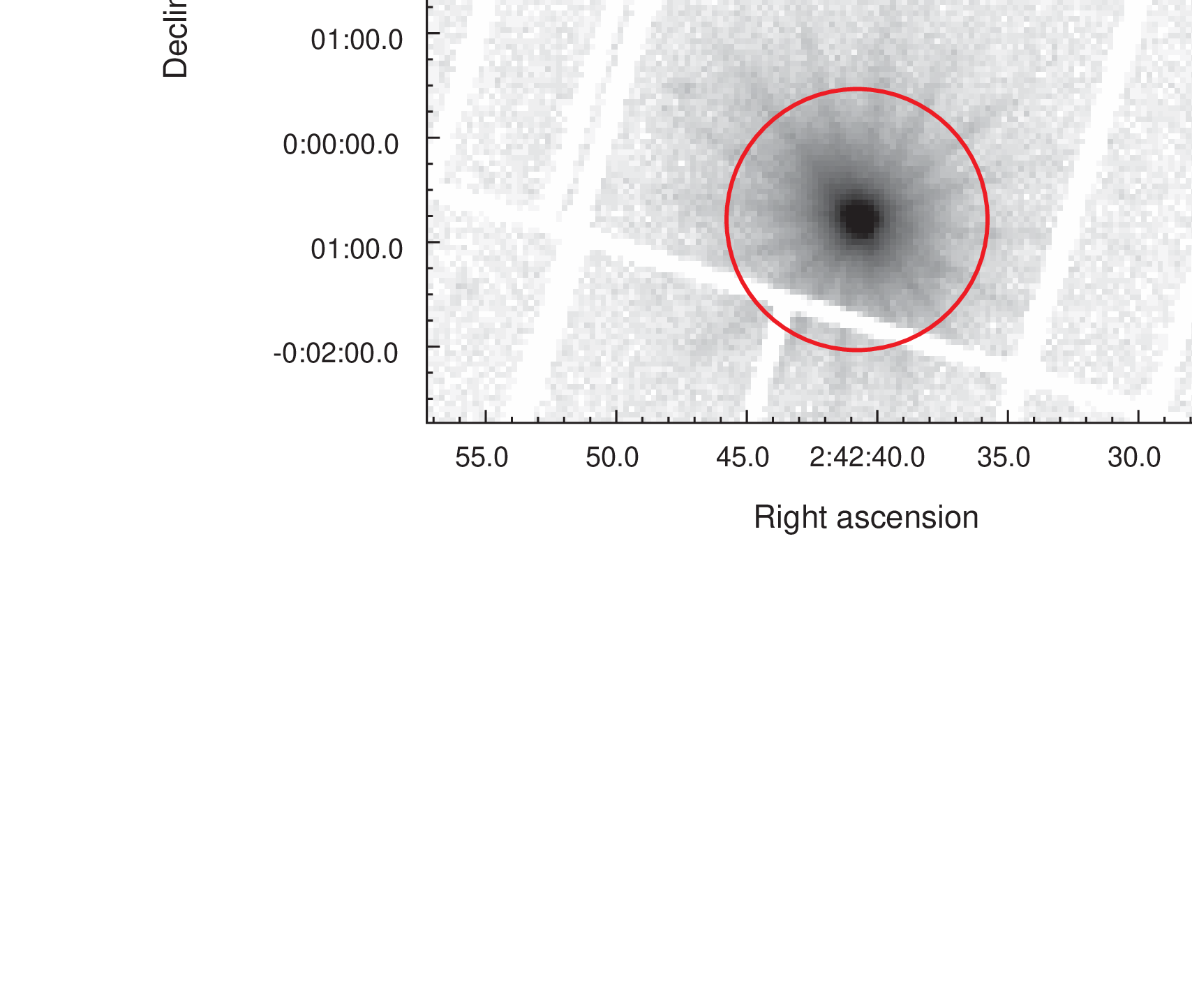}}}
\vglue3.0cm{\hglue-0.3cm{\includegraphics[angle=0, width=8cm]{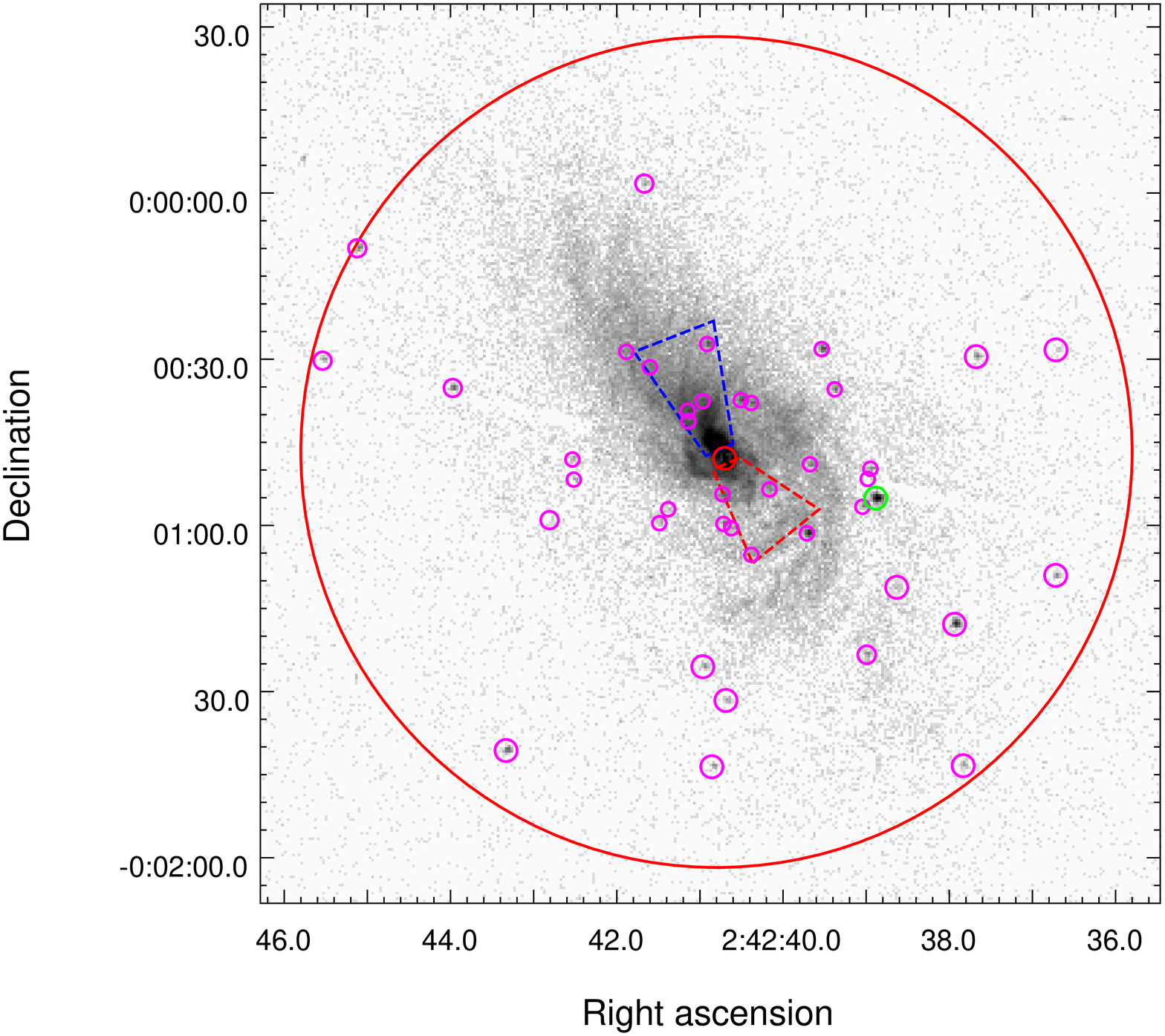}}}
\vspace{-0.05in} \figcaption[]{ 
({\it top}) \nustar{} 3--79~keV FPMA image of \obj showing the
  75\arcsec{} radius source (solid red circle) and polygon background
  (dashed red circle) extraction regions.
({\it middle}) \xmm{} 0.2--10~keV pn image of \obj showing the
  75\arcsec{} radius source (solid red circle) and polygon background
  (dashed red circle) extraction regions. The central point source
  dominates, although there are hints of faint extended emission.
 ({\it bottom}) \chandra{} 0.5--8.0~keV ACIS-S image of \obj showing
  the 75\arcsec{} radius aperture used for \nustar{} and \xmm{} (large
  solid red circle). The nucleus is denoted by the small 2\arcsec{}
  radius (solid red) aperture and is strongly piled up. The obvious
  off-nuclear point sources (denoted by 2--3\arcsec{} radius magenta
  circles) and diffuse emission between 2--75\arcsec{} were extracted
  separately. The rough positions of the radio jet (blue dashed
  region) and counter jet (red dashed region) are shown
  \citep{Wilson1987}. The brightest off-nuclear point source in the
  \chandra{} image (green circle) is not present after 2000-12-04 and
  thus has been excluded from analysis. A background was extracted
  from a source free region on the same chip, outside this figure.
\label{fig:images}}
\vspace{0.5cm}
\end{figure}

NGC\,1068 is the only well-detected source in the \nustar{} FOV (see
Figure~\ref{fig:images}) and appears unresolved. The campaign was
spread over three observations (60002030002, 60002030004, and
60002030006) comprising 123.9\,ks in FPMA and 123.7\,ks in
FPMB. NGC\,1068 appeared as a point source for \nustar{}
(Figure~\ref{fig:images}), and thus spectral products and lightcurves
from both the nucleus and the galaxy emission (diffuse + point
sources) were extracted using 75\arcsec{} radius apertures
(corresponding to $\approx$81\% encircled energy fraction), with
backgrounds estimated from blank regions free of contaminating point
sources on the same detector (see Figure~\ref{fig:images}). We find
that \obj is securely detected up to $\approx55$\,keV at 3$\sigma$
confidence with \nustar{}, and has a maximum signal-to-noise of
$\approx$26 around the peak of Fe K$\alpha$.

We also generated a model of the expected background for each FPM
within our adopted aperture using {\sc nuskybgd} \citep{Wik2014}. {\sc
  nuskybgd} uses several user-defined background regions to sample all
four detectors in each FPM, which it simultaneously fits in order to
model the spectral and angular dependencies for several background
components (e.g., instrumental, focused, and unfocused), before
ultimately generating the expected background within the adopted
aperture. We confirmed the similarity, particularly at high energies
where the background makes a significant contribution, between the
local and model backgrounds to a few percent.  Ultimately we adopted
the local background for simplicity.

\begin{figure*}[t]
\vspace{-0.0in}
\hglue-0.5cm{\includegraphics[angle=270, width=18cm]{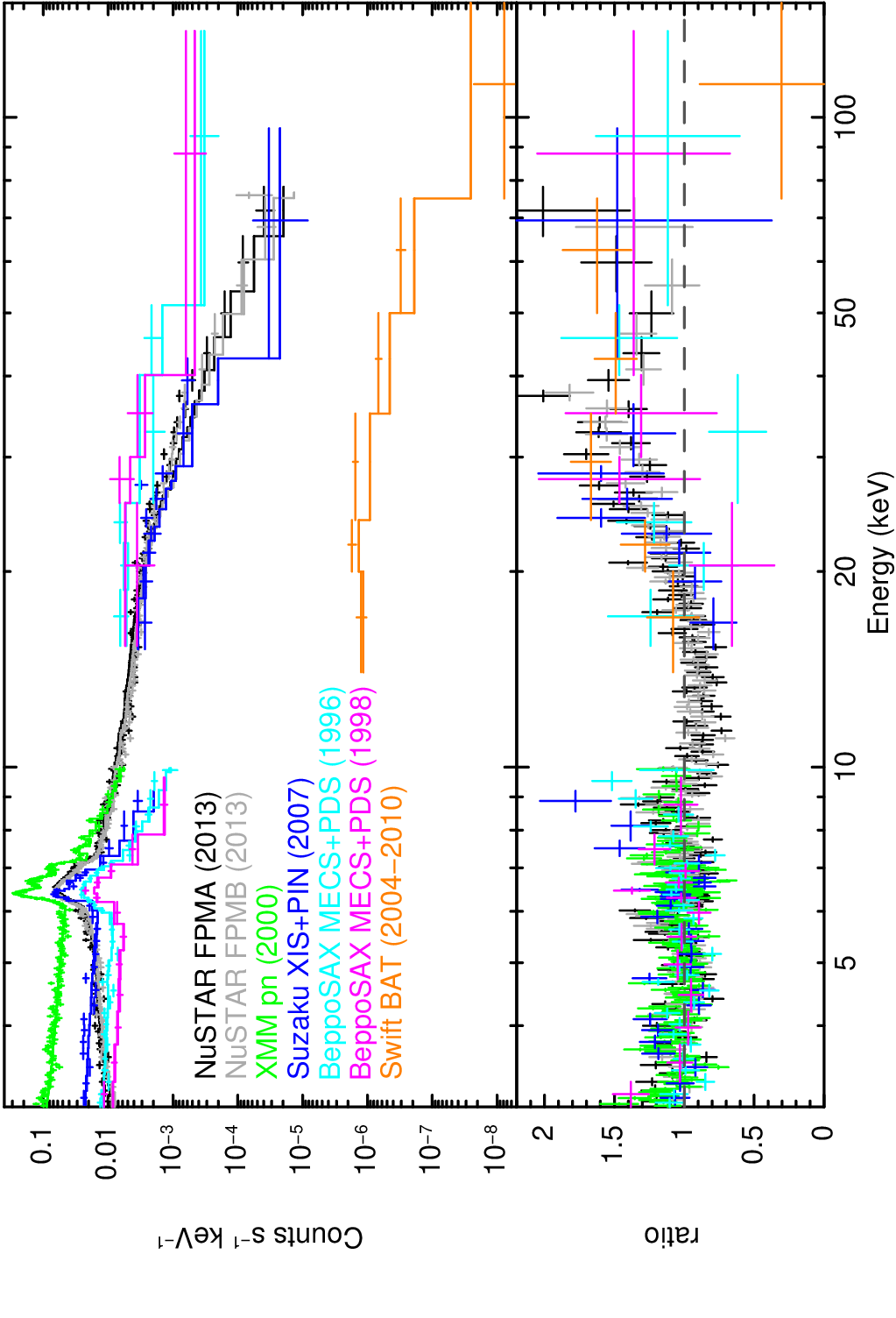}}
\vspace{0.0cm} \figcaption[]{Comparison of the \nustar{} FPMA/FPMB
  (black/grey) and \xmm{} $pn$ (green) spectra with other past
  observations of \obj from \suzaku{} XIS$+$PIN (blue), \bepposax{}
  MECS$+$PDS 1996 (magenta), \bepposax{} MECS$+$PDS 1998 (cyan), and
  \swift{} BAT (orange), all modeled with the best-fitted
  two-reflector model M04a. The top panel shows the observed spectra
  while the bottom panel shows the data-to-model ratios for each
  spectrum. There is good overall consistency between the various
  datasets once known normalization offsets are accounted for, with
  only a few marginally discrepant points seen from the 1996 \bepposax{}
  data. It is clear from the bottom panel that the model provides a
  poor fit to the data near the Compton reflection hump, with the data
  peaking at $\sim$30\,keV while the {\tt pexrav} model peaks at
  $\sim$20\,keV. There are some additional residuals around
  10--15\,keV indicating the curvature of the reflection is more
  severe than the model predicts, as well as around the Fe/Ni line
  region ($\approx6$--8\,keV), suggesting that a few Gaussian lines
  are insufficient for modeling the complex Fe/Ni emission.
\label{fig:M04_fix_allhard}}
\vspace{0.5cm} 
\end{figure*}

Custom position-dependent response matrices and ancillary response
files were generated for the spectra of each module, which provide
nominal vignetting and PSF aperture corrections. In total, we have
$\approx$27,300 and $\approx$26,100 counts between 3--79~keV in FPMA
and FPMB, respectively. As can be seen in
Figure~\ref{fig:M04_fix_allhard}, the FPMA and FPMB spectra are in
excellent agreement, and thus we merged them into a single spectrum
using exposure-weighting for convenience and we use this spectrum for
all fitting and plotting purposes (from Figure~\ref{fig:M04} on).
With respect to the \xmm{} EPIC $pn$ instrument, preliminary results
suggest \nustar{} normalization offsets of 1.11$\pm$0.01; this value
is fully consistent with other \nustar{}/\xmm{} cross-calibration
studies \citep[e.g.,][]{Walton2013, Walton2014}.  Due to the high
signal-to-noise of the \nustar{} data, we also find that we need to
apply a $\approx+40$~eV energy offset (i.e., $\approx$1 spectral bin)
to bring the intrinsic Fe K$\alpha$ line energy (6.4007 keV) into
agreement with the established redshift of \obj and the high
significance line energy determined by the \chandra{} High Energy
Transmission Grating \citep[HETG][]{Canizares2000}; the reason for the
offset is not known, however, its value is within the nominal
calibration precision of \nustar{} and somewhat smaller offsets have
been observed in other sources.

\subsection{\xmm{}}\label{sec:data_xmm}

NGC\,1068 was observed on 2000 July 29--30 with \xmm{} using the EPIC
pn and MOS1/MOS2 instruments \citep{Jansen2001}, which provide
respective angular resolutions of $\approx$5--6\arcsec{} FWHM and
14--15\arcsec{} HPD over the 0.15--12 keV X-ray band,
respectively. Although the energy resolution of the EPIC detectors
(FWHM$\approx$45--150~eV between 0.4--8~keV) is poorer than the HETG,
the difference narrows to a factor of only $\approx$5 by 6--8~keV, and
the three EPIC detectors have substantially larger effective areas
compared to {\it Chandra}. This improvement in counting statistics
allows us to obtain novel constraints on the nuclear spectrum of \obj
compared to the HETG spectra alone.

The \xmm{} observation of \obj was split into two segments made using
the Medium filter in Large Window mode (48\,ms frame-time) for the pn,
in Full Frame mode (1.4\,s frame-time) for MOS1, and in Small Window
mode (0.3\,s frame-time, 110\arcsec$\times$110\arcsec{} FOV) for
MOS2. Given the X-ray flux from the AGN, these options mean that MOS1
will be slightly piled-up while MOS2 will not sample the entire
75\arcsec{} radius extraction region (missing some extended emission
and requiring a larger PSF correction). To limit systematic
uncertainties, we opted to only extract counts for the pn instrument,
which comprises 60\% of the total \xmm{} collecting area (i.e.,
MOS1+MOS2+pn). These are effectively the same conclusions
arrived at by \citet[][hereafter M04]{Matt2004}.

We processed both data sets using SAS (v13.0.0) and selected only
single and double events with quality flag$=$0. The events files were
filtered to exclude background flares selected from time ranges where
the 10--12 keV count rates in the pn camera exceeded 0.3 c/s. The
remaining good exposures are 32.8\,ks for the first observation and
28.7\,ks for the second observation, with $\approx$760,000 counts
between 0.2--10.0~keV.

Source spectra were extracted from a circular region of 75\arcsec{}
radius (corresponding to an $\approx$93.5\% encircled energy fraction)
centered on the nucleus, to match the \nustar{} extraction
region. Background photons were selected from a source-free region of
equal area on the same chip as the source. We constructed response
matrices and ancillary response files using the tasks {\sc rmfgen} and
{\sc arfgen} for each observation. Given that the two observations are
consecutive and constant within their errors, we merged the spectral
products using exposure-weighting. As mentioned previously, we base
all of the normalization offsets relative to the \xmm{} pn, which thus
has a value of 1.00. Additionally, we find we need to apply a
$\approx+15$~eV energy offset (i.e., $\approx$1 response bin) to bring
the intrinsic Fe K$\alpha$ line energy (6.4007 keV) into agreement
with the established redshift of \obj and the high significance line
energy determined by the \chandra{} HETG.

%
%
%

\subsection{{\it Chandra} HETG and ACIS-S}\label{sec:data_hetgs}

NGC\,1068 was observed on multiple occasions with \chandra{} with both
the ACIS-S detector \citep{Garmire2003} by itself and the HETG placed
in front of the ACIS-S. By itself, ACIS-S has a angular resolution of
$<0\farcs5$ FWHM and $\la0\farcs7$ HPD, and a spectral resolution of
FWHM$\approx$110--180~eV between 0.4--8~keV. The HETG consists of two
different grating assemblies, the High Energy Grating (HEG) and the
Medium Energy Grating (MEG), which provide relatively high spectral
resolution (HEG: 0.0007--0.154~eV; MEG: 0.0004--0.063~eV) over the
entire \chandra{} bandpass (HEG: 0.8--10~keV; MEG: 0.4--8~keV). The
gratings operate simultaneously, with the MEG/HEG dispersing a
fraction of the incident photons from the two outer/inner High
Resolution Mirror Assembly (HRMA) shells, respectively, along
dispersion axes offset by 10\degr{}, such that they form a narrow
X-shaped pattern on the ACIS-S detector. Roughly half of the photons
that are not absorbed by the grating pass through undispersed
(preferentially the higher-energy photons) and comprise the HETG 0th
order image on ACIS-S with the standard spectral resolution.

All of the \chandra{} data were reduced following standard procedures
using the {\sc ciao} (v4.5) software package and associated
calibration files (CALDB v4.5.5.1). The data were reprocessed to apply
updated calibration modifications, remove pixel randomization, apply
the energy-dependent sub-pixel event-repositioning (EDSER) techniques,
and correct for charge transfer inefficiency (CTI). The data were
filtered for standard {\it ASCA} grade selection, exclusion of bad
columns and pixels, and intervals of excessively high background (none
were found).  Analysis was performed on reprocessed \chandra{} data,
primarily using {\sc ciao}, but also with custom software.

The 1st-order HETG spectral products were extracted using standard
CIAO tools using a HEG/MEG mask with a full-width of 4\arcsec{} in the
cross-dispersion direction centered on the \obj nucleus; anything
smaller than this will suffer from significant energy-dependent PSF
losses. The intrinsic ACIS-S energy resolution allows to separate the
overlapping orders of the dispersed spectra. The plus and minus sides
were combined to yield single HEG and MEG 1st-order spectra. All of
the HETG data were combined after double-checking that they did not
vary to within errors; obsID 332 appears to have a modestly higher
count rate, but this difference is largely below 2 keV and does not
materially affect the combined $>$2 keV spectra. In total, we have
438.7\,ks of HETG-resolution nuclear spectra available for spectral
fitting (see Table~\ref{tab:data_xray} for details), with
$\approx$12,500 HEG counts between 0.8--10.0~keV and $\approx$34,000
MEG counts between 0.4--8.0~keV. We consider these to be the least
contaminated AGN spectra available below 10~keV (hereafter, simply the
HETG ``AGN'' spectra). The normalization offset between the HEG and
MEG was found to be 1.03$\pm$0.07, while the offsets with respect to
the pn were 1.05$\pm$0.06 and 1.09$\pm$0.06, respectively. This is
consistent with the cross-calibration finding in \citet{Marshall2012}
and \citet{Tsujimoto2011}.

In principle, we have a similar amount of HETG 0th-order data, in
addition to 47.7\,ks of normal ACIS-S data that could be used to model
the extranuclear contamination which strongly affects the lower-energy
\nustar{} and \xmm{} spectra. However the calibration of the HETG
0th-order still remains somewhat uncertain above $\sim5$ keV
(M. Nowak, private communication), which we consider critical for
extrapolating into the \nustar{} band. Thus we chose to model the
contamination spectra solely using ACIS-S obsid 344. These data were
taken with the nominal 3.2s frame time, such that the nucleus is
heavily piled-up ($\sim40$\%) within 1--2\arcsec{}. We therefore
excluded the inner 2\arcsec{} from the contamination analysis and
consider the 2--75\arcsec{} ACIS-S spectrum to be predominantly
emission from the host galaxy (hereafter ``host''), although we must
consider contributions from the broad wings of the PSF (which only
contribute $\approx$5--10\% beyond 2\arcsec{} based on PSF
simulations) and any truly extended Compton reflection and scattered
components from the intrinsic nuclear emission (hereafter ``extended
AGN''). We replaced readout streak events from the nucleus with an
estimate of the background using the {\sc acisreadcorr} tool.

We note that the brightest off-nuclear point source,
CXOU\,J024238.9$-$000055.15, which lies 30\arcsec{} to the southwest
of the nucleus in the 344 observation, provides $\sim$20\% of the host
contamination counts above 4 keV. Notably, it is not present after
2000-12-04 in either the \chandra{} or \swift{} observations, the latter
of which is simultaneous with the \nustar{} observation (see
$\S$\ref{sec:data_swift}). Thus we excluded this point source from our
assessment of the host contamination contribution to the \nustar{}
spectra.  The source is present and distinct during the \xmm{}
observations, and is found to comprise $\approx1.5$\% of the $>4$ keV
counts (see also the extended discussion in M04). For simplicity, we
account for its presence in the \xmm{} spectrum of \obj using an
additional normalization adjustment between \xmm{} and \nustar{}.

To assess host contamination, initially we extracted ACIS-S spectra of
obvious point sources and diffuse emission separately, as shown in
Figure~\ref{fig:images}, using {\sc specextract}. For the point
sources, we used 2--3\arcsec{} extraction radii, depending on whether
they are strong or weak and whether they reside within strong diffuse
emission, while for the diffuse emission we extracted everything else
between 2--75\arcsec{}. In total, we found $\approx$6300 and
$\approx$93450 0.4--8.0~keV counts for the off-nuclear point source
and diffuse components, respectively. A local background region was
extracted from an adjacent region 100\arcsec{} in radius
$\approx3\farcm5$ northwest of \obj. Ultimately, to simplify the
contamination model and improve statistics, we also extracted a total
contamination spectrum of all emission within 2--75\arcsec{}. The
normalization offset between the pn and ACIS-S was 1.04$\pm$0.04,
which is consistent with the values found by \citet{Nevalainen2010}
and \citet{Tsujimoto2011}.

\subsection{\bepposax}\label{sec:data_bepposax}

NGC\,1068 was observed by \bepposax{} on 1996 December 30 and on 1998
January 11 with the Low Energy Concentrator Spectrometer (LECS), the
three Medium Energy Concentrator Spectrometers (MECS), and Phoswich
Detector System (PDS). We use only the MECS and PDS here. 

The MECS contains three identical gas scintillation proportional
counters, with an angular resolution of $\approx0\farcm7$ FWHM and
$\approx2\farcm5$ HPD, and a spectral resolution of
FWHM$\approx$200--600~eV between 1.3--10~keV. MECS1 failed a few
months after launch and thus only MECS2 and MECS3 data are available
for the 1998 observation. The MECS event files were screened adopting
standard pipeline selection parameters. Spectra were extracted from
3\arcmin{} radii apertures and the spectra from individual units were
combined after renormalizing to the MECS1 energy-PI
relation. Background spectra were obtained using appropriate blank-sky
files from the same region as the source extraction. The resulting
MECS spectra have $\approx$5900 counts between 3--10~keV in 100.8\,ks
of good exposure for the first observation and $\approx$1550 counts in
37.3\,ks for the second observation. We find that the MECS
normalization is systematically offset from the pn by a factor of
1.12$\pm$0.02 in the 3--7\,keV band and thus by a factor of
1.02$\pm$0.02 with respect to \nustar{} in the same band.

The PDS has no imaging capability, but does have sensitivity between
15--220~keV and can potentially provide some constraints above the
\nustar{} band. The PDS data were calibrated and cleaned using the
SAXDAS software within HEASoft, adopting the 'fixed Rise Time
threshold' method for background rejection. The PDS lightcurves are
known to show spikes on timescales of fractions of second to a few
seconds, with most counts from the spikes typically falling below 30
keV. We screened the PDS data for these spikes following the method
suggested in the NFI user
guide,\footnote{http://heasarc.nasa.gov/docs/sax/abc/saxabc/saxabc.html}
arriving at $\approx$16,600$\pm$3010 counts between 15--220~keV in
62.5\,ks of good exposure for the first observation and
$\approx$4720$\pm$1560 counts in 17.7\,ks for the second
observation. The PDS spectra were logarithmically rebinned between
15--220 keV into 18 channels, although we cut the spectrum at 140\,keV
due to poor statistics. With the data quality/binning, it is difficult
to appreciate the presence of a bump at 30 keV. The PDS normalization
is known to be low by $\approx$20--30\% \citep{Grandi1997, Matt1997}
compared to the MECS, which we accounted for by using a fixed
normalization constant of 0.7$\pm$0.1 when modeling the data with
respect to \nustar{}.

We note that the statistics for the second \bepposax{} observation are
poorer, with many of the channels statistically consistent with zero.

\subsection{\suzaku}\label{sec:data_suzaku}

The \suzaku{} observatory observed \obj with the X-ray Imaging
Spectrometer (XIS) and Hard X-ray Detector (HXD) PIN instruments on
2007 February 10. Our reduction follows the recommendations of the
\suzaku{} Data Reduction
Guide.\footnote{http://heasarc.gsfc.nasa.gov/docs/suzaku/analysis/} 

For the XIS, we generated cleaned event files for each operational
detector (XIS0, XIS1, and XIS3) and both editing modes (3x3 and 5x5)
using the \suzaku{} {\sc aepipeline} with the latest calibration, as
well as the associated screening criteria files in HEASoft. Using {\sc
  xselect}, source spectra were extracted using a 260\arcsec{} radius
aperture, while background spectra were extracted from remaining
regions free of any obvious contaminating point sources. Responses
were generated for each detector using the {\sc xisresp} script with a
medium resolution. The spectra for the front-illuminated detectors
XIS0 and XIS3 were consistent, and were subsequently combined using
{\sc addascaspec}; for simplicity, we adopt this composite spectrum to
represent the XIS. We obtained $\approx$33,300 counts with a good
exposure of 61.5~ks. We find that the XIS spectrum is systematically
offset from the pn by a factor of 1.17$\pm$0.02, which is slightly
(i.e, $<3\sigma$) above the expected normalization offset of
$1.10\pm0.01$ assessed by \citet[][]{Tsujimoto2011}.

Similar to the PDS, the PIN has poor angular resolution
(0\fdg56$\times$0\fdg56 FOV) but does have sensitivity between
15--70~keV and thus provides another point of comparison with
\nustar{}. We reprocessed the unfiltered event files following the
data reduction guide to obtain $\approx$15,500 counts with a good
exposure of 39.0~ks. No significant detection was found in the
GSO. Since the HXD is a collimating instrument, estimating the
background requires separate consideration of the non X-ray
instrumental background (NXB) and cosmic X-ray background (CXB), which
comprise $\approx$89\% of the total counts. We used the response and
NXB files provided by the \suzaku{}
team,\footnote{http://www.astro.isas.ac.jp/suzaku/analysis/hxd/}
adopting the model D `tuned' background. Spectral products were
generated using the {\sc hxdpinxbpi} tool, which extracts a composite
background using the aforementioned NXB and a simulated contribution
from the expected CXB following \citet{Boldt1987}. We find the PIN
normalization to be systematically offset from \nustar{} by a factor
of 1.2$\pm$0.05, which is consistent with the current
cross-calibration uncertainty (K. Madsen et al., submitted).

\begin{figure*}[t]
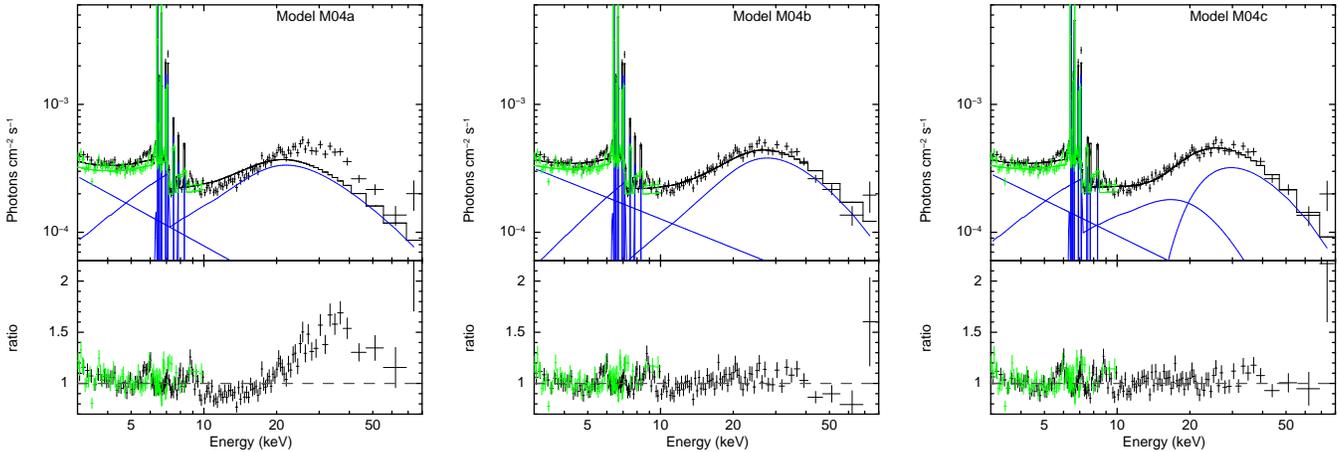

\vspace{-0.0in}
\hglue-0.5cm{\includegraphics[angle=270, width=6cm]{Model_Matt2004_fix.ps}}
\hglue-0.0cm{\includegraphics[angle=270, width=6cm]{Model_Matt2004_free.ps}}
\hglue-0.0cm{\includegraphics[angle=270, width=6cm]{Model_Matt2004_leaky.ps}}
\vspace{0.0cm} \figcaption[]{Comparison of the \nustar{} (black) and
  \xmm{} pn (green) spectra of \obj, modeled with the best-fitted
  two-reflector model (blue) where several variables are fixed ({\it
    left}; model M04a), fit as free parameters ({\it middle}; model
  M04b), and with the addition of a leaky, absorbed transmission
  component ({\it right}; model M04c). The top panel shows the
  observed spectra while the bottom panel shows the data-to-model
  ratios for each spectrum. The overall fits with the parameters free
  and the addition of the transmission component
  \citep[e.g.,][]{George2000} are better, with most of the residuals
  confined to the complex Fe/Ni line region.
\label{fig:M04}}
\vspace{0.5cm} 
\end{figure*}

\subsection{\swift{}}\label{sec:data_swift}

The \swift{} observatory observed \obj with the X-ray Telescope (XRT;
7\arcsec{} FWHM, 20\arcsec{} HPD) for $\approx$2\,ks simultaneous with
\nustar{} on 2012 December 19. The processed data were retrieved from
the \swift{} archive, and analysis was performed using {\sc
  ftools}. With $\approx$1200 counts between 0.5--10~keV in a
75\arcsec{} aperture, the \swift{} exposure is not long enough to
provide additional constraints beyond those already obtained with
\nustar{}, \xmm{}, and \chandra{}. However, it does serve to determine
if any transient point sources strongly contributed to the $<$10~keV
\nustar{} spectra of \obj. To this end, we generated a 0.5--10~keV
image with {\sc xselect}, which is consistent with the \chandra{}
images from 2008 to within the limits of the \swift{} XRT angular
resolution and does not show any new strong off-nuclear point
sources. We find the XRT 3--10\, keV composite spectrum is consistent
with the other instruments aside from its normalization, which is
systematically offset from the pn by a factor of 1.12$\pm$0.25; the
large error bar is due to the fact that the observation only has 64
counts in the 3--10\,keV band. This offset is fully consistent with
those found by \citet{Tsujimoto2011}.

Since November 2004, the Burst Alert Telescope (BAT) onboard \swift{}
\citep{Gehrels2004} has been monitoring the hard X-ray sky
(14--195\,keV) and can potentially provide some constraints above the
\nustar{} band.  \swift{} BAT uses a 5200 cm$^{2}$ coded-aperture mask
above an array of 32,768 CdZnTe detectors to produce a wide field of
view of 1.4 steradian of the sky and an effective resolution of
$\approx$20\arcmin{} (FWHM) in stacked mosaicked maps. Based on the
lack of variability (see $\S$\ref{sec:variability}), we used
the stacked 70-month spectrum, which is extracted from the central
pixel \citep[2\farcm7;][]{Baumgartner2013} associated with the BAT
counterpart, to assess nature of the emission. The
background-subtracted spectrum contains $\approx460$ counts in the
14--195~keV band.  We find the BAT normalization to be systematically
offset from \nustar{} by a factor of 0.75$\pm$0.05, which is
consistent with the current cross-calibration uncertainty (K. Madsen et al.,
in preparation).

\section{X-ray Variability Constraints}\label{sec:variability}

Depending on the location and structure of the obscuration in \obj, it
may be possible to observe temporal variations in one or more of its
spectral components on short or long timescales. Notably,
there have been previous claims of low-significance variability from
the warm reflection component between the \bepposax{} and \xmm{}
observations \citep{Guainazzi2000, Matt2004}. 

\begin{figure}[t]
\vspace{-0.0in}
\hglue-0.2cm{\includegraphics[angle=270, width=8cm]{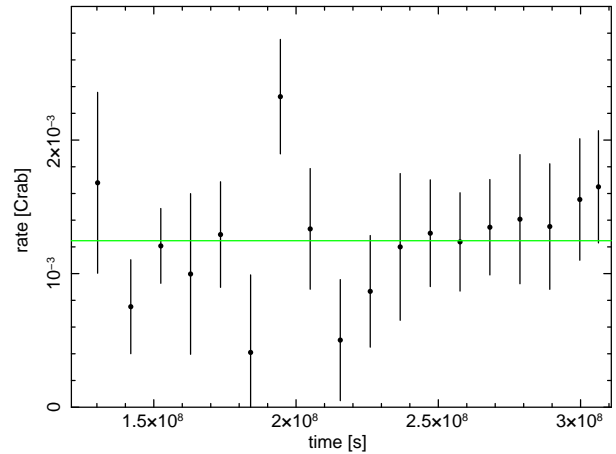}}
\vspace{0.0cm} \figcaption[]{\swift{} BAT 70-month 14-195\,keV light
  curve spanning 2004-2010, in bins of $\approx$0.5\,yr. The green
  line denotes the best-fitted constant model ($\chi_{\nu}^{2}=0.95$
  for $\nu=17$), indicating that \obj shows no significant hard X-ray
  variability over this time span.
\label{fig:BAT_lc}}
\vspace{0.5cm} 
\end{figure}

As shown in Table~\ref{tab:data_xray}, we find reasonable consistency
between the count rates extracted from all instruments where \obj was
observed more than once, with differences always less than 3$\sigma$
based on counting statistics. These constraints imply there is no
strong continuum variability below $10$\,keV over periods of 10--15
years. Since \swift{} BAT continuously observes the sky, a new
snapshot image can be produced every $\sim$1 week for persistent
high-energy X-ray sources due to the wide field of view and large sky
coverage. To study long-term variability of \obj
(SWIFT\_J0242.6$+$0000) above 10\,keV, we use the publicly available
70-month (9.3 Ms) lightcurves from \swift{} BAT
\citep{Baumgartner2013}, which span 2004--2010. The wide energy range
of \swift{} BAT allows us to test any underlying energy dependence of
the lightcurve, assessing lightcurves in eight non-overlapping energy
bands: 14--20, 20--24, 24--35, 35--50, 50--75, 75--100, 100--150,
150--195 keV. The cumulative 14--195\,keV lightcurve, binned in
half-year intervals due to the limited statistics, is shown in
Figure~\ref{fig:BAT_lc} and is formally constant to within errors
($\chi_{\nu}^{2}=0.95$ for $\nu=17$ degrees of freedom). Variability
limits in the individual bands are consistent with the full band
results, but generally are less constraining due to limited
statistics.

\begin{figure}[t]
\vspace{-0.0in}
\hglue-0.2cm{\includegraphics[angle=270, width=8cm]{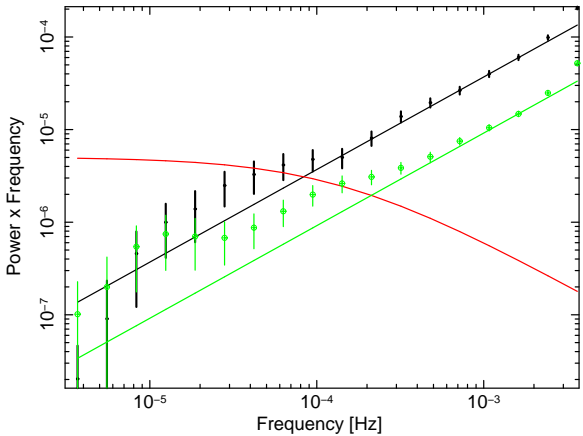}}
\vspace{0.0cm} \figcaption[]{ Power spectra of the combined 30--79 keV
  band lightcurves from \nustar{} for \obj.  The source and background
  power spectra are plotted in black and green symbols, respectively.
  The black and green lines denote the power spectra expected for
  Poisson noise only for each case.  The solid red curve represents
  the power spectrum of the direct continuum for an AGN of the same
  mass and accretion rate as \obj. The high-energy lightcurves are
  roughly consistent (at $\approx2\sigma$) with simple Poisson noise,
  although there could be additional low-frequency noise that affects
  both the source and background variability. Furthermore, backgrounds
  extracted from the other \nustar{} FPM detectors produced similar
  shapes. Thus we conclude that the variability constraints are
  significantly below the expected value for a transmitted AGN
  continuum.
\label{fig:NU_freq}}
\vspace{0.5cm} 
\end{figure}

To investigate short-term variability, we applied the
Kolmogorov-Smirnov (K-S) test to individual observations, finding all
observations to be constant in count rate with 3$\sigma$
confidence. We searched for additional hints of short-term variability
taking advantage of the high throughput of \nustar{} above
10\,keV. The timescales covered by these light curves
($\sim$1--200\,ks) can only reveal rapid fluctuations, such as those
expected from the intrinsic powerlaw emission. Therefore, any
variability seen in this range would be indicative of a transmitted
powerlaw component \citep[e.g.][]{Markowitz2003, McHardy2004,
  McHardy2005, McHardy2006, Markowitz2007}. In
Figure~\ref{fig:NU_freq}, we constructed power spectra from the
high-energy \nustar{} lightcurves for \obj and a typical background
region, which we compare to the expected power spectra for pure
Poisson noise and for the expected variability of a pure transmitted
component, as observed in unobscured AGN of similar mass and accretion
rates. To produce this, we extracted 30--79 keV counts from our
nominal source region and a background region of equal area on the
same detector using {\sc xselect}. We constructed lightcurves in 100 s
equally spaced bins, retaining only those which had exposure ratios
over 90\%. Note that the nature of \nustar{}'s orbit means that for
the given sky location we will have 2 ks gaps in the lightcurves every
6 ks. Moreover, since \nustar{} observed \obj in three distinct
segments, we have larger gaps in between the observations. To mitigate
these potential sources of aliasing, we calculated power spectra using
the Mexican-hat filtering method described in \citet{Arevalo2012},
which is largely unaffected by gaps in the lightcurves. Finally, we
normalized the power as the variance divided by the square of the
average count rate. As can be seen in Figure~\ref{fig:NU_freq}, the
power spectrum detected from \obj is fully consistent with Poisson
noise at better than 2$\sigma$.\footnote{It is important to stress
  here that the convolution kernel is broad in frequency, such that
  nearby power density spectral points will be correlated. Thus the
  fact that several consecutive points are above the PN level does not
  make the detection of variability more significant.} Thus, if there
is any transmitted component leaking through, it does not comprise the
bulk of the $>10$\,keV flux.

We conclude that if there has been any variability from \obj in the
past $\approx$15 years, it has been at a level comparable to either
the cross-calibration uncertainties between instruments or the
statistical uncertainty in the data and that the short-term behaviour
as measured by the \nustar{} lightcurves is not consistent with a
transmitted powerlaw component dominating the flux above 10\,keV.

\section{X-ray Spectral Constraints}\label{sec:spectra}

We begin by comparing the high-quality combined \nustar{} and \xmm{}
spectra to those from several past satellites to demonstrate the
dramatic improvement in data quality. We compare all of these to a few
common previously used models, which can eventually fit the data
relatively well when pushed to extreme values. Following this, we
develop a more realistic approach to quantify the non-negligible
contamination from extranuclear emission and then model the AGN
components using a few common models such as {\tt pexmon}
\citep{Nandra2007}, {\tt MYTorus} \citep{Murphy2009, Yaqoob2012}, and
      {\tt torus} \citep{Brightman2011}.

Unless stated otherwise, modeling was performed with \xspec v12.9.0
(Arnaud 1996), and quoted uncertainties on spectral parameters are
90\% confidence limits for a single parameter of interest, and
spectral fitting is performed through $\chi^{2}$ minimization. Neutral
absorption is treated with the {\tt tbabs} absorption code
\citep{Wilms2000}, with appropriate solar abundances ({\tt wilm}) and
cross sections \citep[{\tt vern};][]{Verner1996}.

Throughout our analysis, we assume there is no angular dependence of
the nuclear emission spectral shape (such that all scatterers see the
same photon index) and we neglect any accretion disk reflection
component \citep[e.g.,][]{Ross2005, Dauser2013, Garcia2014} when
modeling the obscured nuclear radiation, which is justified due to the
inclination and dominance of scattering and absorption from distant
material.

Finally, we note that \xspec has considerable difficulty arriving at
the best-fit solution when dealing with large numbers of free
parameters, such as we have in \obj associated with the considerable
line emission. Thus, to mitigate this in cases in which we fit
individual emission lines separately, we individually fitted the line
centers, redshifts, widths, and heights of the Gaussian lines over
small portions of the spectrum above a local powerlaw continuum, and
then froze each line at its best-fit values. We then fit the relative
contributions from the continuum and fluorescent line models.

\subsection{Comparison to Previous Models}\label{sec:previous}

As mentioned in $\S$\ref{sec:intro}, \obj has been successfully
modeled in the past above $\approx$3--4~keV with a double reflector
comprised of both neutral ``cold'' \citep[{\tt pexrav with
    $R=-1$};][]{Magdziarz1995} and ionized ``warm'' ({\tt cutoffpo})
Compton-scattered components, plus a few Gaussian emission lines to
model the strong Fe and Ni emission \citep[hereafter model ``M04a'',
  since it was adopted from M04; see also similar models from
]{Matt1997, Guainazzi2000}. We therefore began by fitting this model
(see Table~\ref{tab:specmodels}) to the \nustar{}, \xmm{},
\bepposax{}, \suzaku{}, and \swift{} BAT spectra above 3 keV.

\begin{deluxetable*}{ll}
\tabletypesize{\scriptsize}
\tablewidth{0pt}
\tablecaption{X-ray Spectral Fitting Models\label{tab:specmodels}}
\tablehead{
\colhead{Model} &
\colhead{{\sc xspec} Components}
}
\tableheadfrac{0.05}
\startdata
\hline
\multicolumn{2}{c}{Total} \\
\hline
M04a/b & {\scriptsize {\tt tbabs*(pexrav+cutoffpo+zgauss(Fe$_{\rm neutral}$, Fe$_{\rm ionized}$, Ni$_{\rm neutral}$))}} \\
M04c   & {\scriptsize {\tt tbabs*(MYTZ*cutoffpo+cutoffpo+pexrav+zgauss(Fe$_{\rm neutral}$, Fe$_{\rm ionized}$, Ni$_{\rm neutral}$))}} \\
\hline
\multicolumn{2}{c}{Nucleus Only} \\
\hline
P      & {\scriptsize {\tt tbabs(tbabs(MYTZ*cutoffpo+C$_{\rm RRC}$+C$_{\rm RL}$+pow+zedge(Ni)*gsmooth(pexmon)+zgauss(Ni K$\beta$)))}} \\
M1     & {\scriptsize {\tt tbabs(tbabs(MYTZ*cutoffpo+C$_{\rm RRC}$+C$_{\rm RL}$+pow+highecut*zedge(Ni)*MYTS+gsmooth*MYTL+zgauss(Ni$_{\rm neutral}$)))}} \\
M2     & {\scriptsize {\tt tbabs(tbabs(MYTZ*cutoffpo+C$_{\rm RRC}$+C$_{\rm RL}$+pow+highecut*zedge(Ni)*MYTS(0\degr{}, 90\degr{})+gsmooth*MYTL(0\degr{}, 90\degr{})+zgauss(Ni$_{\rm neutral}$)))}} \\
T      & {\scriptsize {\tt tbabs(tbabs(C$_{\rm RRC}$+C$_{\rm RL}$+pow+highecut*gsmooth(torus)))}}\\
\hline
\multicolumn{2}{c}{Host Only} \\
\hline
P      & {\scriptsize {\tt tbabs(pcfabs(C$_{\rm RRC}$+C$_{\rm RL}$+pow+zedge(Ni)*gsmooth(pexmon)+zgauss(Ni K$\beta$)))}} \\
M1     & {\scriptsize {\tt tbabs(pcfabs(C$_{\rm RRC}$+C$_{\rm RL}$+pow+highecut*zedge(Ni)*MYTS+gsmooth*MYTL+zgauss(Ni$_{\rm neutral}$)))}} \\
M2     & {\scriptsize {\tt tbabs(pcfabs(C$_{\rm RRC}$+C$_{\rm RL}$+pow+highecut*zedge(Ni)*MYTS(0\degr{}, 90\degr{})+gsmooth*MYTL(0\degr{}, 90\degr{})+zgauss(Ni$_{\rm neutral}$)))}} \\
T      & {\scriptsize {\tt tbabs(pcfabs(C$_{\rm RRC}$+C$_{\rm RL}$+pow+highecut*gsmooth(torus)))}}
\enddata
\tablecomments{
We denote Fe$_{\rm neutral}$ to signify the modeling of neutral Fe
K$\alpha$ (6.40\,keV) and K$\beta$ (7.07\,keV) transitions, while we
use Ni$_{\rm neutral}$ for the modeling of neutral Ni K$\alpha$
(7.47\,keV) and K$\beta$ (8.23\,keV) lines. We denote Fe$_{\rm ionized}$
to signify the modeling of ionized Fe K$\alpha$ H-like (6.97\,keV),
He-like (6.69\,keV), and Be-like (6.57\,keV) transitions.
We denote C$_{\rm RRC}$ to signify the modeling of the radiative
recombination continuum, which is modeled by a 0.3\,keV {\tt bremss}
component.
We denote C$_{\rm RL}$ to signify the modeling of the radiative
recombination line emission, which is comprised of numerous
transitions from a variety of elements. We adopted line species,
energies and strengths consistent with those reported in K14 (which
includes Fe$_{\rm ionized}$), as well as Ni He-like K$\alpha$
(7.83\,keV).  For the ACIS-S host spectrum, we model only a subset of
lines comprising just the strongest handful of K14 lines.
}
\end{deluxetable*}


We initially fixed most of the parameters to the values found by
M04\footnote{Note that in {\tt pexrav}, the inclination angle is
  specified in units of $\cos{90-\theta_{\rm inc}}$, such that a value
  of 0.88 is equivalent to 63\degr{}.  } (e.g., $\Gamma=2.04$, $Z_{\rm
  Fe} = 2.4Z_{\rm \odot,Fe}$, $\theta_{\rm inc}=63$\degr{}, $E_{\rm
  cut}=500$~keV), varying only the component normalizations and the
redshifts of the emission lines. The normalizations were coupled
between the different instruments while the redshifts differed for
each instrument to account for the aforementioned linear energy
offsets.  The redshifts of the cold reflector and neutral lines
(K$\alpha$, K$\beta$) were tied and allowed to vary as one parameter,
while the redshifts of the ionized lines were tied and allowed to vary
as another parameter. The best fit of this dual-reflector model, M04a,
yielded a reduced $\chi^{2}_{\nu}=1.40$ for $\nu=1785$. As can be seen
in Figure~\ref{fig:M04_fix_allhard}, the fit has strong residuals near
the Compton reflection hump due to a discrepancy between the peak of
the reflection hump in the data ($\sim$30\,keV) and the one from the
{\tt pexrav} model ($\sim$20\,keV). We also see residuals around
10--15\,keV, implying that there is stronger curvature in the actual
reflection spectrum than has been modeled, as well as around the Fe/Ni
line complex, suggesting that the Gaussians are not sufficient to
describe the line complexity observed.

The bottom panel of Figure~\ref{fig:M04_fix_allhard} shows the
data-to-model ratios for several past hard X-ray missions compared
against the best-fitted fixed-$\Gamma$ two-reflector model. As noted
in $\S$\ref{sec:variability}, there have been previous claims of
low-level variability in the warm reflection component
\citep{Guainazzi2000, Matt2004}. After accounting for known
cross-calibration offsets, we find that the \nustar{}, \xmm{} pn,
\suzaku XIS, and \bepposax{} MECS spectra in the 3--5\,keV range,
where the warm reflector should dominate, are consistent within their
statistical uncertainties based on powerlaw fits to this range; this
applies to the 3--10\,keV range overall as well. Uncertainties in the
normalization offsets between instruments, and hence flux differences,
above 10\,keV are considerably larger, making it more difficult to
assess potential variability. Nonetheless, after accounting for known
cross-calibration offsets, we find that the \nustar{}, \suzaku{} PIN,
1998 \bepposax{} PDS, and \swift{} BAT spectra above 10\,keV are
likewise consistent within their statistical uncertainties. The 1996
\bepposax{} PDS spectra, which lack the pronounced residuals around
30~keV that we observe from the other hard X-ray spectra, differ from
the rest at marginal significance (2.5$\sigma$) and in fact appear to
be relatively well-fitted by the fixed \citet{Matt2004} model
($\chi^{2}_{\nu}=1.43$ for $\nu=57$ by itself; perhaps this is no
surprise since the model is based on these data). Here, it is
important to remember that the \bepposax{} PDS, \suzaku{} PIN, and
\swift{} BAT spectra are all strongly background-dominated (see
$\S$\ref{sec:data_bepposax}-\ref{sec:data_swift}), and minor
variations in background levels (e.g., due to minor flares or how the
data are screened) can potentially lead to large variations in the
source spectra. The fact that we see an overall consistency in the
spectral shape of the residuals, aside from the one discrepant point
in the 1996 \bepposax{} PDS spectra around 30 keV, demonstrates that
there has been no strong variability detected over at least the past
$\approx$15 years.

\begin{deluxetable*}{lcrrrrrrrrrrr}
\tabletypesize{\scriptsize}
\tablewidth{0pt}
\tablecaption{Model Spectral Fit Parameters\label{tab:fits}}
\tablehead{
\colhead{(1)} & 
\colhead{(2)} & 
\colhead{(3)} & 
\colhead{(4)} & 
\colhead{(5)} & 
\colhead{(6)} & 
\colhead{(7)} & 
\colhead{(8)} & 
\colhead{(9)} & 
\colhead{(10)} & 
\colhead{(11)} & 
\colhead{(12)} & 
\colhead{(13)} \\
\colhead{Model} & 
\colhead{Spectra} & 
\colhead{Range} & 
\colhead{$\Gamma$} &
\colhead{$N_{\rm H}$} &
\colhead{$E_{\rm c}$} &
\colhead{$\theta_{\rm inc}$} &
\colhead{$\theta_{\rm open}$} &
\colhead{$Z_{\rm Fe}$} &
\colhead{S/L ratio} &
\colhead{$\log{F_{\rm X, cold}}$} &
\colhead{$\log{F_{\rm X, warm}}$} &
\colhead{$\chi^{2}_{\nu}$ ($\nu$)}
}
\tableheadfrac{0.05}
\startdata
\hline
\multicolumn{13}{c}{Total (Previous Models)} \\
\hline
M04a  & XN &  3--79 & $2.04$                 & 10 & $500$             & 63    & \nodata & $2.4$\rlap{$^{*}$}        & \nodata & $-11.70^{+0.01}_{-0.01}$ & $-11.67^{+0.01}_{-0.01}$ & 1.61 (1234) \\
M04b  & XN &  3--79 & $1.76^{+0.04}_{-0.09}$ & 10 & $108^{+19}_{-18}$ & $70^{\ ---}_{-7}$ & \nodata & $6.8^{+0.4}_{-0.4}$\rlap{$^{*}$} & \nodata & $-11.86^{+0.02}_{-0.02}$ & $-11.57^{+0.01}_{-0.01}$ & 1.20 (1230) \\
M04c  & XN &  3--79 & $1.92^{+0.05}_{-0.06}$ & 10$^{\ ---}_{-0.05}$ & $22^{+24}_{-9}$   & 63    & \nodata & $2.4$\rlap{$^{*}$}       & \nodata & $-11.74^{+0.01}_{-0.01}$ & $-11.64^{+0.01}_{-0.01}$ & 1.22 (1227) \\
\hline
\multicolumn{13}{c}{Nucleus Only} \\
\hline
P               & H & 0.5--9 & 2.46$^{\ ---}_{-0.24}$ & 10  & 500 & 85 & \nodata & $4.5^{+1.1}_{-0.6}$ & \nodata & -11.80$^{+0.02}_{-0.02}$ & -11.75$^{+0.04}_{-0.04}$ & 1.60 (1472) \\
P               & H &   2--9 & 1.15$^{+0.32}_{\ ---}$ & 10  & 500 & 85 & \nodata & $5.1^{+3.7}_{-0.9}$ & \nodata & -11.77$^{+0.12}_{-0.17}$ & -11.79$^{+0.14}_{-0.22}$ & 0.66  (319) \\
M1              & H & 0.5--9 & 1.40$^{+0.12}_{\ ---}$ & 10  & 500 & 90 & 60  & \nodata & 0.42$^{+0.12}_{-0.08}$ & -11.82$^{+0.08}_{-0.05}$ & -11.81$^{+0.02}_{-0.09}$ & 1.64 (1472) \\
M1              & H &   2--9 & 1.40$^{+0.16}_{\ ---}$ & 10  & 500 & 90 & 60  & \nodata & 0.46$^{+0.13}_{-0.08}$ & -11.73$^{+0.02}_{-0.07}$ & -11.87$^{+0.10}_{-0.09}$ & 0.72 (319) \\
M2              & H & 0.5--9 & 2.60$^{\ ---}_{-0.19}$ & 10  & 500 & 0,90 & 60  & \nodata & 0.67$^{+0.09}_{-0.09}$ & -11.54$^{+0.02}_{-0.02}$ & -12.36$^{+0.14}_{-0.16}$ & 1.62 (1471) \\
M2              & H &   2--9 & 1.52$^{+0.01}_{-0.08}$ & 10  & 500 & 0,90 & 60  & \nodata & 0.64$^{+0.13}_{-0.11}$ & -11.87$^{+0.14}_{-0.11}$ & -11.76$^{+0.03}_{-0.16}$ & 0.74 (318) \\
T               & H & 0.5--9 & 1.30$^{+0.09}_{-0.05}$ & 10  & \nodata & 87  & 67$^{+12}_{-15}$ & \nodata & \nodata & -11.78$^{+0.03}_{-0.05}$   & -11.76$^{+0.02}_{-0.02}$ & 1.65 (1472) \\
T               & H &   2--9 & 1.14$^{+0.33}_{\ ---}$ & 10  & \nodata & 87  & 67$^{+11}_{-17}$ & \nodata & \nodata & -11.82$^{+0.30}_{-0.09}$   & -11.69$^{+0.06}_{-0.09}$ & 0.73 (319) \\
\hline
\multicolumn{13}{c}{Host Only} \\
\hline
P               & A & 0.5--9 & 2.49$^{\ ---}_{-0.19}$ & \nodata & 500 & 85 & \nodata & $33^{+40}_{-13}$ & \nodata & -12.47$^{+0.08}_{-0.15}$ & -12.09$^{+0.04}_{-0.01}$ & 1.42 (163) \\
P               & A &   2--9 & 2.49$^{\ ---}_{-0.40}$ & \nodata & 500 & 85 & \nodata & $100^{\ ---}_{-60}$ & \nodata & -12.45$^{+0.11}_{-0.22}$ & -12.11$^{+0.09}_{-0.03}$ & 0.81 (73) \\
M1              & A & 0.5--9 & 2.55$^{\ ---}_{-0.06}$ & 10  & 500 & 90 & 60  & \nodata & 2.46$^{+3.49}_{-1.01}$ & -12.37$^{+0.11}_{-0.16}$ & -12.11$^{+0.03}_{-0.03}$ & 1.44 (163) \\
M1              & A &   2--9 & 2.60$^{\ ---}_{-0.42}$ & 10  & 500 & 90 & 60  & \nodata & 2.27$^{\ ---}_{-0.90}$ & -12.35$^{+0.07}_{-0.38}$ & -12.12$^{+0.10}_{-0.03}$ & 0.89 (73) \\
M2              & A & 0.5--9 & 2.56$^{\ ---}_{-0.06}$ & 10  & 500 & 0,90 & 60  & \nodata & 2.25$^{+2.65}_{-0.90}$ & -12.36$^{+0.10}_{-0.15}$ & -12.11$^{+0.04}_{-0.03}$ & 1.44 (162) \\
M2              & A &   2--9 & 2.60$^{\ ---}_{-0.35}$ & 10  & 500 & 0,90 & 60  & \nodata & 2.70$^{+5.20}_{-1.09}$ & -12.34$^{+0.10}_{-0.33}$ & -12.13$^{+0.10}_{-0.03}$ & 0.87 (72) \\
T               & A & 0.5--9 & 2.61$^{+0.06}_{-0.06}$ & 10  & \nodata & 87  & 26$^{+13}_{-26}$ & \nodata & \nodata & -12.24$^{+0.07}_{-0.08}$   & -12.17$^{+0.03}_{-0.03}$ & 1.49 (163) \\
T               & A &   2--9 & 2.91$^{\ ---}_{-0.39}$ & 10  & \nodata & 87  & 26$^{+17}_{-26}$ & \nodata & \nodata & -12.17$^{+0.07}_{-0.10}$   & -12.25$^{+0.10}_{-0.17}$ & 0.94 (73) \\
\hline
\multicolumn{13}{c}{Total (Nucleus$+$Host Models)} \\
\hline
Pa  & HAXNB & 2--195 & 1.57$^{+0.02}_{-0.02}$ & 10  & 500                  & 85               & \nodata & $5.0^{+0.3}_{-0.3}$ & \nodata & $-11.87^{+0.02}_{-0.02}$ & $-11.90^{+0.02}_{-0.02}$ & 1.34 (1666) \\
    &       &        &                        &     &                      &                  & \nodata &        & \nodata & $-12.40^{+0.05}_{-0.05}$ & $-12.60^{+0.10}_{-0.13}$ & \\
M1a & HAXNB & 2--195 & 1.40$^{+0.09}_{\ ---}$ & 10  & 500                  & 90 & \nodata & \nodata & 1 & $-14.00^{+0.30}_{\ ---}$ & $-11.67^{+0.01}_{-0.01}$ & 3.78 (1666) \\
    &       &        &                        &     &                      &                  & \nodata & \nodata & & $-14.00^{+0.30}_{\ ---}$ & $-11.92^{+0.03}_{-0.03}$ & \\
M1d & HAXNB & 2--195 & 1.40$^{+0.12}_{\ ---}$ & 9.4$^{\ ---}_{-3.3}$ & 41$^{+5}_{-4}$ & 78$^{+3}_{-4}$ & \nodata & \nodata & $3.8^{+0.5}_{-0.8}$ & $-12.44^{+0.04}_{-0.04}$ & $-11.78^{+0.01}_{-0.01}$ & 1.31 (1662) \\
    &       &        &                        &     &                 &             & \nodata & \nodata & & $-13.33^{+0.14}_{-0.22}$ & $-11.99^{+0.02}_{-0.02}$ & \\
M1g & HAXNB & 2--195 & 1.40$^{+0.34}_{\ ---}$ & 10  & 34$^{+58}_{-4}$ & 80.7$^{+6.5}_{-3.4}$ & & \nodata \nodata & $3.3^{+0.8}_{-0.5}$ & $-12.05^{+0.01}_{-0.01}$ & $-11.97^{+0.02}_{-0.02}$ & 1.29 (1660) \\
    &       &        &                        &     &                 & 88.3$^{\ ---}_{-21.1}$ & \nodata & \nodata &  $1.5^{+0.3}_{-0.2}$ & $-12.01^{+0.03}_{-0.03}$ & $-12.83^{+0.17}_{-0.29}$ & \\
%
%
M2a & HAXNB & 2--195 & 2.29$^{+0.04}_{-0.02}$ & 10  & 500                  & 90 & \nodata & \nodata & 1 & $-11.87^{+0.01}_{-0.02}$ & $-12.10^{+0.02}_{-0.02}$ & 1.83 (1666) \\
    &       &        &                        &     &                      &  0 & \nodata & \nodata &   & $-13.90^{+0.25}_{-0.67}$ &                          & \\
    &       &        &                        &     &                      &  0 & \nodata & \nodata &    & $-14.00^{+0.30}_{\ ---}$ & $-12.08^{+0.02}_{-0.02}$ & \\
M2d & HAXNB & 2--195 & 2.10$^{+0.06}_{-0.07}$ & 10.0$^{\ ---}_{-0.3}$ & 128$^{+115}_{-44}$ & 90 & \nodata & \nodata & 1.0$^{+0.1}_{-0.1}$ & $-11.81^{+0.02}_{-0.02}$ & $-12.34^{+0.05}_{-0.05}$ & 1.14 (1666) \\
    &       &        &                        & 0.14$^{+0.01}_{-0.01}$ &                &  0 & \nodata & \nodata & & $-12.92^{+0.03}_{-0.03}$ &                          & \\
    &       &        &                        & 5.0$^{+4.2}_{-1.9}$ &                      &  0 & \nodata & \nodata & & $-12.36^{+0.04}_{-0.05}$ & $-12.20^{+0.03}_{-0.03}$ & \\
Ta  & HAXNB & 2--195 & 1.96$^{+0.05}_{-0.04}$ & 10  & 500               & 87 & 64$^{+3}_{-2}$ & 1          & \nodata & $-11.95^{+0.02}_{-0.03}$ & $-11.86^{+0.02}_{-0.02}$ & 1.61 (1667) \\
    &       &        &                        & 10  &                      &                  &     &                      & \nodata & $-14.00^{+0.30}_{\ ---}$ & $-11.92^{+0.03}_{-0.03}$ & \\
Tc  & HAXNB & 2--195 & 2.13$^{+0.04}_{-0.06}$ & 6.3$^{+0.6}_{-0.8}$ & 500               & 87$^{\ ---}_{-12}$ & 69$^{+4}_{-3}$ & 1          & \nodata & $-11.96^{+0.02}_{-0.03}$ & $-11.87^{+0.02}_{-0.02}$ & 1.57 (1663) \\
    &       &        &                        & 10$^{\ ---}_{-6.6}$ &                   &                  &     &                      & \nodata & $-14.00^{+0.30}_{\ ---}$ & $-11.92^{+0.03}_{-0.03}$ & \\
\enddata
\tablecomments{
{\it Column 1}: Model used. Model name beginning with: ``M04'' denote variations of M04 models; ``P'' denote variations of {\tt pexmon} models; ``M1'' denote variations of coupled {\tt MYTorus} models; ``M2'' denote variations of decoupled {\tt MYTorous} models; ``T'' denote variations of {\tt Torus} models. See $\S$\ref{sec:previous} and $\S$\ref{sec:new} for details. When multiple rows are listed, the first
one or two rows represent the nucleus model while the second or third
row represents the host model.
{\it Column 2}: Spectra fit, where X$=$\xmm{} pn, A$=$\chandra{} ACIS-S, H$=$\chandra{} HEG$+$MEG, N$=$\nustar{}, and B$=$\swift{} BAT.
{\it Column 3}: Energy range fit, in keV.
{\it Column 4}: Photon index of the primary transmitted powerlaw
continuum. Note that some reported limits are poorly constrained since
the allowed ranges for $\Gamma$ are confined to between 1.1--2.5 for the
{\tt pexmon} model, between 1.4--2.6 for the {\tt MYTorus} model,
and between 1--3 for the {\tt torus} model.
{\it Column 5}: Neutral hydrogen column density of the obscuring
torus/clouds, in units of $10^{24}$\,cm$^{-2}$.
{\it Column 6}: Energy of the exponential cutoff rollover of primary
transmitted powerlaw continuum, in keV.
{\it Column 7}: Inclination angle with respect to a face-on geometry,
in degrees. Note that some reported limits are poorly constrained
since the allowed ranges for $\theta_{\rm inc}$ are confined to
0\degr{}--72\degr{} for the M04 ({\it pexrav}) model,
0\degr{}--85\degr{} for the {\tt pexmon} model, and 18\fdg2-87\fdg1
for the {\tt torus} model.
{\it Column 8}: Torus opening angle, in degrees. This parameter is not
meaningful for the M04 and {\tt pexmon} models, is fixed at 60\degr{}
for the {\tt MYTorus} model, and is confined to 25\fdg8-84\fdg3 for
the {\tt torus} model.
{\it Column 9}: Fe abundance with respect to our adopted value of
$Z_{\odot, Fe}$.  The overall abundance of metals (not including Fe)
is assumed to be solar ($Z_{\odot}$). Note that entries denoted by
$^{*}$ are for {\tt pexrav}, where the Fe abundance is driving the
peak of the Compton reflection hump to higher energy and has no effect
on the Fe line emission, which is modeled with Gaussians.
{\it Column 10}: Ratio of the scattered and line components of {\tt
  MYTorus}. This can crudely be interpreted as an Fe abundance with
respect our adopted value of $Z_{\odot, Fe}$, although care should be
taken since the correspondence is non-trivial and hence only approximate
\citep{Yaqoob2012}. 
{\it Columns 11--12}: Logarithms of the 2--10\,keV fluxes of the cold
and warm reflection components, respectively, in units of
erg\,s$^{-1}$\,cm$^{-2}$.
{\it Column 13}: Reduced $\chi^{2}_{\nu}$ and degrees of freedom for
given model.
Values with no quoted errors were fixed at their specified values.
}
\end{deluxetable*}


We note that the $\chi^{2}$ residuals are dominated by the \xmm{} pn
and \nustar{} spectra, and thus, for clarity, we opt to use only the
\xmm{} pn and combined FPMA/FPMB \nustar{} spectra to represent the
global spectrum of \obj hereafter. To this end, we plot the unfolded
\xmm{} pn and composite \nustar{} spectra along with the various
components that comprise the M04a model again in the left panel of
Figure~\ref{fig:M04}, as well as the data-to-model residuals.  This
fit yielded a reduced $\chi^{2}_{\nu}=1.61$ for $\nu=1234$. The
continuum parameter values and errors are listed in
Table~\ref{tab:fits}, while the normalizations of the various lines
are given in Table~\ref{tab:lines}. For \nustar{}, the redshifts for
the neutral and ionized lines were -0.0065$^{+0.0005}_{-0.0006}$ and
0.0081$^{+0.0012}_{-0.0010}$, respectively, while for \xmm{} they were
0.0015$^{+0.0003}_{-0.0008}$ and 0.0026$^{+0.0012}_{-0.0013}$,
respectively.


Allowing the powerlaw index, high-energy cutoff, and Fe abundance and
inclination angle of the reflector to vary, hereafter model ``M04b'',
improves the fit substantially, with a reduced $\chi^{2}_{\nu}=1.20$
for $\nu=1230$. As shown in Figure~\ref{fig:M04}, most of the
residuals are now due to the Fe/Ni line complex with only very mild
residuals seen from the Compton hump above 10 keV. The emission line
parameters remained more or less constant, while the best fitted
values of the other parameters are $\Gamma=1.76^{+0.04}_{-0.09}$,
$\theta_{\rm inc}=70^{+20}_{-7}$, $E_{\rm c}=108^{+19}_{-18}$\,keV,
and $Z_{\rm Fe}=6.8\pm0.4$. Parameter values and errors for model M04b
are listed in Table~\ref{tab:fits}.


Another possibility that could explain the spectrum is if the direct
continuum is partially punching through above 20--30\,keV, often
called the ``leaky'' torus model, hereafter model ``M04c''. Given the
high column density needed to produce flux only above $\sim$30\,keV,
we need to account properly for the effects of Compton absorption,
which we do through the use of the multiplicative transmission
component from the {\tt MYTorus} set of models \citep[hereafter {\tt
    MYTZ} to denote ``zeroth-order'' component;][]{Murphy2009} and a
cutoff power law ({\tt cutoffpl}); see also $\S$\ref{sec:new}. For
this direct component, we tie the values of the intrinsic continuum
slope, cutoff energy, and redshift to those of the scattered
components, which were left to vary. The normalizations for the three
continuum components were free to vary as well.  The inclination angle
and Fe abundance of the {\tt pexrav} component of M04c were fixed at
63\degr{} and 2.4, respectively, as in M04a, to limit the number of
free parameters. This model yields a reduced $\chi^{2}_{\nu}=1.22$ for
$\nu=1227$, with most of the residuals due to the Fe line complex and
only very mild residuals around the Compton hump above 10 keV. As
before, the emission line parameters remained more or less constant,
while the best fitted values of the other parameters are a photon
index of $\Gamma=1.92^{+0.05}_{-0.06}$, an exponential cutoff rollover
energy of $E_{\rm c}=22^{+24}_{-9}$\,keV, a column density for the
absorbed transmission component of $N_{\rm
  H}>9.95\times10^{24}$~cm$^{-2}$, and normalizations of $A_{\rm
  trans}=67.8$$^{+2.5}_{-1.8}$ \pnorm{}, $A_{\rm
  warm}=(7.9\pm0.5)\times10^{-4}$ \pnorm{}, and $A_{\rm
  cold}=(1.1^{+2.4}_{-1.8})\times10^{-2}$ \pnorm{}.  Parameter values
for model M04c are listed in Table~\ref{tab:fits}.

Clearly the best-fitted M04a model fails to provide an adequate
description of the \nustar{} data, while both of the alternative
models, M04b and M04c appear to yield more reasonable fits. The Fe
abundance constraints from M04b are substantially super-solar, which
is consistent with past constraints on \obj
\citep[e.g.,][]{Kinkhabwala2002, Kraemer1998} as well as some
unobscured AGN \citep[e.g.,][]{Fabian2009, Fabian2013, Parker2014},
although such results are not necessarily definitive. The
overabundance is at least partially driven by the need to fit the
30\,keV bump with a much deeper iron edge. Thus model M04b remains
potentially viable. Model M04c provides an equally acceptable fit,
although it requires that the transmitted component dominates above
20\,keV with rather unusual best-fit parameters. For instance, the
cutoff energy implies a unrealistically low corona temperature
\citep[e.g.,][]{Petrucci2001}, while the ratio of
transmitted-to-scattered normalizations is abnormally high ($\approx
6000$). As such, this scenario seems unlikely on its own and can be
further ruled out by the variability constraints presented in
$\S$\ref{sec:data_nustar} and $\S$\ref{sec:data_swift}, which imply
that \obj is more or less constant over all of the timescales which we
have measured.

One might be tempted to stop here, having modeled the global \nustar{}
and \xmm{} spectra to a reasonably acceptable level. However, higher
spectral and angular resolution data from \chandra{} exist, allowing us to
remove potential host contamination and thus probe the nature of the
scattering medium in more detail. Additionally, a critical drawback of
the {\tt pexrav} model, for instance, is that it models a simple
slab-like geometry for the Compton scatterer assuming an infinite
column density, which almost certainly fails to adequately describe
the true physical situation (e.g., a smooth or clumpy torus) present
in \obj.  To this end, we also explore a variety of models which adopt
more realistic geometrical scenarios for AGN scattering in section
$\S$\ref{sec:new}.

\begin{deluxetable}{lrrrrr}
\tabletypesize{\scriptsize}
\tablewidth{0pt}
\tablecaption{Fe and Ni Line Fluxes (with Model M04b)\label{tab:lines}}
\tablehead{
\colhead{} & 
\multicolumn{2}{c}{\xmm{}} & &
\multicolumn{2}{c}{\chandra{}} \\ \cline{2-3} \cline{5-6}\\
\colhead{Line}  &
\colhead{M04} &
\colhead{this work} & &
\colhead{HETG} &
\colhead{ACIS-S} \\
\colhead{}  &
\colhead{$<$40\arcsec} &
\colhead{$<$75\arcsec} & &
\colhead{$<$2\arcsec} &
\colhead{2\arcsec-75\arcsec}
}
\tableheadfrac{0.05}
\startdata
Fe neutral K$\alpha$         &      44.3          & 47.4$^{+1.9}_{-2.2}$ & &    38.9$^{+3.8}_{-3.8}$ &    17.5$^{+3.3}_{-3.3}$ \\
Fe neutral K$\alpha$ CS      &       8.7          &  3.8$^{+1.5}_{-2.2}$ & &     4.2$^{+3.6}_{-2.2}$ & $<$ 1.5 \\
Fe neutral K$\beta$          &       9.1          &  8.9$^{+1.1}_{-1.5}$ & &     4.3$^{+3.1}_{-1.9}$ & $<$ 5.2 \\
Ni neutral K$\alpha$         &       5.6          &  5.8$^{+1.8}_{-0.9}$ & & $<$ 7.3                 & $<$ 8.8 \\
Ni neutral K$\beta$          &       3.2          &  3.1$^{+0.9}_{-1.3}$ & & $<$19.8                 & $<$16.8 \\
Fe Be-like 6.57\,keV &    7.6\rlap{*} &  8.0$^{+1.5}_{-2.1}$ & &     6.3$^{+2.1}_{-2.5}$ &     3.9$^{+2.9}_{-2.9}$ \\
Fe He-like 6.69\,keV &   22.8\rlap{*} & 27.8$^{+1.0}_{-2.9}$ & &    12.8$^{+3.9}_{-2.3}$ &     6.1$^{+2.4}_{-2.4}$ \\
Fe H-like  6.97\,keV &    7.1\rlap{*} &  8.2$^{+0.8}_{-2.5}$ & &     7.7$^{+1.5}_{-5.8}$ & $<$ 6.1 \\
Ni He-like 7.83\,keV &    2.7\rlap{*} &  3.9$^{+1.1}_{-1.1}$ & & $<$10.2                 & $<$10.4
\enddata
\tablecomments{
{\it Column 1}: Primary Fe and Ni lines measured in M04 and
here using the \xmm{} and \chandra{} datasets. The continuum was fit in
all cases with the M04b model for consistency.
{\it Columns 2-5}: Normalizations of the best-fitted {\tt zgauss}
components to each line, in units of $10^{-6}$
photons\,s$^{-1}$\,cm$^{-2}$. The components denoted by *'s were
mistakenly listed in Table 3 of M04 with values a factor of 10 higher
than intended (G. Matt, private communication); they have been
corrected here for clarity. The difference between \xmm{} pn values is
at least partially due to differences in encircled energy fractions
(87\% vs. 93\%) between the extraction regions.  }
\end{deluxetable}

\subsection{Detailed Spectral Modeling}\label{sec:new}

At this point, it is critical to define which spectral models we
will fit to the data, as there are a variety of models
of Compton-scattered emission which have been used to fit
reflection-dominated spectrum to account for the possible different
geometries of the scattering material. These include

\begin{itemize}
\item {\tt pexmon} ---
this is a modified version of the standard {\tt pexrav} model
\citep{Magdziarz1995} already used in $\S$\ref{sec:previous}, which
self-consistently computes the continuum (based on {\tt pexrav}) as
well as the neutral Fe K$\alpha$, Fe K$\beta$ and Ni K$\alpha$
emission lines \citep[based on Monte Carlo simulations
  by][]{George1991} and the Fe K$\alpha$ Compton shoulder
\citep{Nandra2007}. As with {\tt pexrav}, this model assumes that the
scattering structure has a slab geometry and infinite optical
depth. Moreover, the total and Fe abundances can be adjusted to
account for non-solar values.  The Ni edge is not included in this
model, so we add this as a {\tt zedge} component at the systemic
redshift, the depth of which is tied to the the measured Ni K$\alpha$
flux (a value of $\tau=0.1$ in {\tt zedge} achieved this). Results for
the series of {\tt pexmon} Compton scattering models are detailed
below and summarized in Table~\ref{tab:fits}. We caution that the {\tt
  pexmon} model is limited to photon indices between $\Gamma=$1.1--2.5
and inclination angles $\theta=$0\degr{}--85\degr{}.

\item {\tt MYTorus} --- 
  functions for a smoothly distributed toroidal reprocessor composed
  of gas and dust with finite optical depth and with a fixed 60\degr{}
  opening angle \citep{Murphy2009}. {\tt MYTorus} is comprised of
  three separate spectral components: a transmitted intrinsic
  continuum component ({\tt MYTZ}, incorporated as a multiplicative
  table) which represents the photons along the direct line of sight
  to the nucleus which remain after scattering, and Compton-scattered
  continuum (hereafter {\tt MYTS}) and fluorescent line and Compton
  shoulder (hereafter {\tt MYTL}) components which represent photons
  scattered into our line of sight from a different viewing angle to
  the nucleus (both additive table models). The neutral Fe lines are
  modeled self-consistently with the Compton-scattered component. By
  using multiple scatterers, varying their relative normalizations
  and/or inclination angles with respect to our line of sight,
  dissentangling their column densities, and so forth,
  \citet{Yaqoob2012} demonstrated that one could model a wide range of
  possible geometries surrounding the central engine. The Ni edge is
  not included in this model, so we must add this as a {\tt zedge}
  component at the systemic redshift, the depth of which is tied to
  the measured Ni K$\alpha$ flux (which empirically equates to fixing
  $\tau=0.1$). The model does not allow dynamic fitting for a
  high-energy cutoff, and table models are only computed for a handful
  of fiducial ``termination'' energies ($E_{\tt T}$, which effectively
  is an instant cutoff).\footnote{Below energies of $\approx$20 keV,
    the {\tt MYTorus} models with different termination energies are
    virtually identical, while above this value the lower termination
    energy models have psuedo-exponential cutoffs, the forms of which
    depend modestly on input parameters. Using a sharp termination
    compared to an exponential cutoff should lead to mild differences
    in the shape of the cutoff. Unfortunately, the lack of any
    continuum above the termination energy imposes parameter
    limitations when fitting, e.g., the \swift{} BAT spectrum. While
    there may be merits to the arguments given in the {\tt MYTorus}
    manual against applying a cutoff, we find the alternative, a
    dramatic cutoff, to also be unsatisfactory from a physical
    standpoint.} For expediency, we chose to implement a dynamic
  cutoff separately using the $E_{\it T}=500$\,keV model multipled by
  the {\tt highecut} model with a fixed pivot energy of 10~keV and an
  e-folding energy that is tied to the transmitted powerlaw cutoff
  energy.\footnote{While applying exponential cutoffs outside of {\tt
      MYTorus} is expressly warned against in the {\tt MYTorus}
    manual, we found that this method yielded reasonable consistency
    compared to the various {\tt MYTorus} termination energy models
    over the ranges of parameters we fit, such that constraints on the
    cutoff energies typically were less than a factor of two different
    from the termination energy considered.} It is argued in the {\tt
    MYTorus} manual that applying a high-energy cutoff ruins the
  self-consistency of the {\tt MYTorus}; therefore, once we determined an
  approximate $E_{\rm cut}$ for our best-fit model, we dropped the use of
  {\tt highecut} and replaced the $E_{\rm T}=500$\,keV model with one
  which best approximates $E_{\rm cut}=E_{\rm T}$. Results for the
  series of {\tt MYTorus} Compton scattering models are detailed below
  and summarized in Table~\ref{tab:fits}. We caution that the {\tt
    MYTorus} model is computed only for photon indices between
  $\Gamma=$1.4--2.6, to energies between 0.5--500 keV, and solar
  abundances.

\item {\tt torus} ---
this model describes obscuration by a spherical medium with variable
$N_{\rm H}$ and inclination angle, as well as a variable biconical polar
opening angle \citep[][]{Brightman2011}. {\tt torus} self-consistently
predicts the K$\alpha$ and K$\beta$ fluorescent emission lines and
absorption edges of all the relevant elements. The key advantage of
this model is it can fit a range of opening angles and extends up to
$N_{\rm H}=10^{26}$\,cm${^2}$, but a major drawback is that it does not
  allow the user to separate the transmitted and Compton-scattered
  components. As such, it can only be applied to the nuclear emission
  and is not appropriate to model the host component, which should
  include only the Compton-scattered emission. As with {\tt MYTorus},
  {\tt torus} does not allow dynamic fitting for a high-energy cutoff,
  so we implemented a cutoff using {\tt highecut} in the same manner
  as for {\tt MYTorus}.  Results for the series of {\tt torus} Compton
  scattering models are detailed below and summarized in
  Table~\ref{tab:fits}.  We caution that the {\tt torus} model is
  limited to photon indices between $\Gamma=$1.0--3.0, inclination
  angles $\theta_{\rm inc}=$18\fdg1--87\fdg1, opening angles
  $\theta_{\rm tor}=$25\fdg8-84\fdg3, and solar abundances.

\end{itemize}

\noindent Both {\tt MYTorus} and {\tt torus} provide significant and
distinct improvements over the geometric slab model which manifest in
the spectral shapes of both line and continua. Nonetheless, the reader
should keep in mind that they too only sample a small portion of the
potential parameter space that likely describes real gas distributions
in the vicinity of AGN. As done in $\S$\ref{sec:previous}, we model
the transmitted powerlaw continuum as {\tt MYTZ}*{\tt cutoffpl} where
applicable.

Before we proceed to fitting these more complex models, we note that
while \nustar{} and \xmm{} have large collecting areas and wide energy
coverage, neither is able to spatially separate the spectra of the AGN
from various sources of host contamination (or even extended AGN
emission from point-like AGN emission), nor are they able to spectrally
resolve some line complexes to gain a better understanding of the
physical processes involved \citep[e.g.,K14;]{Kinkhabwala2002}. We
rely on the \chandra{} data for these purposes, allowing us to
construct the most robust model to date for the nuclear and global
spectra of \obj. We use the \chandra{} HETG spectra to model the
point-like nuclear emission from \obj from the inner 2\arcsec{} in
$\S$\ref{sec:nuc} and use the \chandra{} ACIS-S data to model the host
galaxy emission from \obj between 2--75\arcsec{} in
$\S$\ref{sec:host}, both of which are fit between 0.5--9.0
keV. Fitting down to 0.5 keV allows us to constrain the soft-energy
components, which can affect the flux and slope of the ionized
reflector if unaccounted for. We then proceed to fit the combination
of the nuclear and host galaxy emission to the \nustar{}, \xmm{}, and
\chandra{} spectra in $\S$\ref{sec:combined}. We also fit the \swift{}
BAT spectra with this combined spectrum, since the BAT spectrum
provides some additional spectral coverage up to $\approx$200\,keV and
its spectral shape more or less agrees with \nustar{} where the two
spectra overlap (see Figure~\ref{fig:M04_fix_allhard}).

\subsubsection{Point-Like Nuclear Emission}\label{sec:nuc}

\begin{figure}[t]
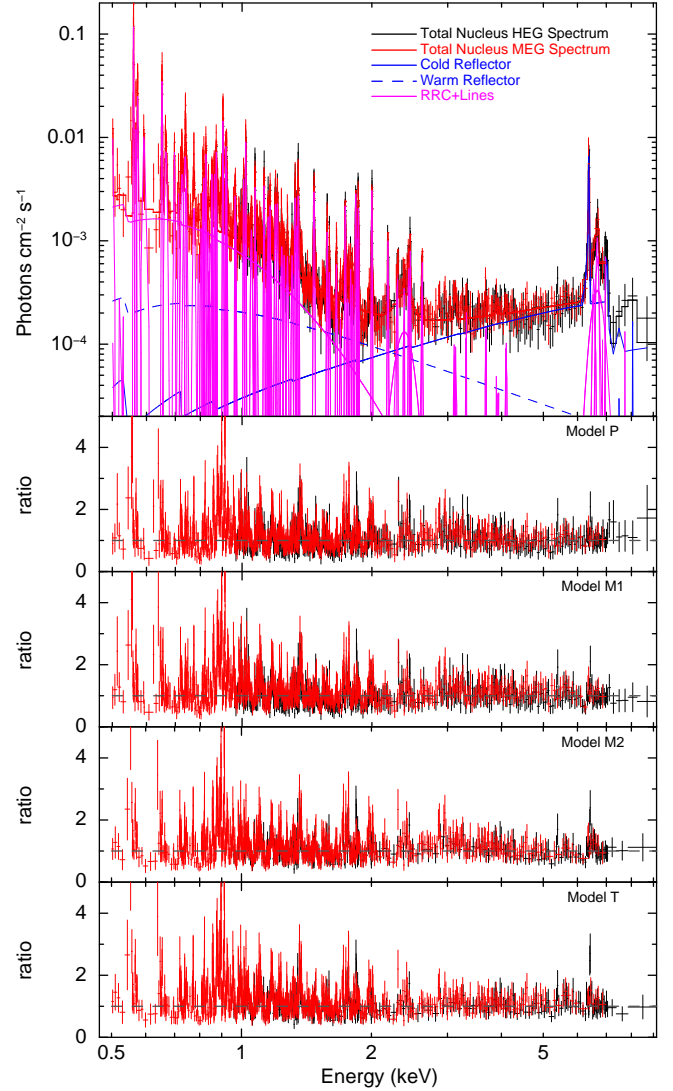

\vspace{-0.0in}
\hglue-0.5cm{\includegraphics[angle=270, width=9.0cm]{Model_P_nuc_hetg.ps}}
\vglue-0.09cm{\hglue-0.43cm{\includegraphics[angle=270, width=8.93cm]{Model_P_nuc_hetg2.ps}}}
\vglue-0.15cm{\hglue-0.43cm{\includegraphics[angle=270, width=8.93cm]{Model_M1_nuc_hetg.ps}}}
\vglue-0.15cm{\hglue-0.43cm{\includegraphics[angle=270, width=8.93cm]{Model_M2_nuc_hetg.ps}}}
\vglue-0.15cm{\hglue-0.43cm{\includegraphics[angle=270, width=8.93cm]{Model_T_nuc_hetg.ps}}}
\vspace{0.0cm} \figcaption[]{ HEG (black) and MEG (red) unfolded
  spectra of nucleus of \obj extracted from a 4\arcsec{} full-width
  mask. The spectra are fitted with four models comprised of
  partially absorbed cold and warm reflectors (blue solid and dashed
  lines, respectively), as well as RRC and line components as
  identified by K14 (magenta solid lines). The top panel shows the fit
  to model P while the rest of the panels show the data-to-model
  ratios for models P, M1, M2, and T, respectively.
\label{fig:hetg_nuc}}
\vspace{0.5cm} 
\end{figure}

The HETG nuclear spectra are shown in Figure~\ref{fig:hetg_nuc}, and
clearly exhibit emission from several different components. To
reproduce the main features of the HETG spectra, we assume that the
nucleus AGN spectrum has a direct transmitted power law ({\tt
  cutoffpl}) with slope $\Gamma_{\rm nuc}$ and cutoff energy $E_{\rm
  c, nuc}$, which is absorbed by a presumably Compton-thick absorber
(e.g., an edge-on torus) with neutral column density $N_{\rm H}$
(modeled as {\tt MYTZ*cutoffpl}). Based on the modeling in
$\S$\ref{sec:previous}, we adopt fixed values of $N_{\rm
  H}=10^{25}$\,cm$^{-2}$ and $\theta_{\rm inc}=90$\degr{} for the
absorber; these quantities are poorly constrained by the $<10$\,keV
data alone due to degeneracies with other spectral components (see
below). Although the transmitted component is not observable below
$\sim$10\,keV (if at all) in Compton-thick AGN, most of the observed
$<$10\,keV features should be indirect products of it. 

We empirically model the soft RRC and RL emission as a bremsstrahlung
component ({\tt bremss}, best-fit $kT_{\rm bremss}=0.31\pm0.03$\,keV
and $A_{\rm bremss}=0.013\pm0.01$ cm$^{-2}$) and $\approx$90 narrow
ionized emission lines ({\tt zgauss}), respectively. The latter are
based on the line identifications from K14 plus Ni He-like
7.83\,keV. For simplicity we adopt a single redshift $z_{\rm ion}$ and
line width $\sigma_{\rm ion}$ (fixed at 0.0035\,keV) for the vast
majority of the ionized lines. The line normalizations are in crude
agreement with K14, although there are differences due to our adopted
widths and redshifts; as these are primarily used so that we can
constrain the RRC {\tt bremss} temperature and normalization, we do
not report the line properties here. Even with all of these
components, significant complex Fe and Si line emission remains (see
Figure~\ref{fig:hetg_nuc}; residuals can also be seen in Figures 1--4
of K14), which we modeled empirically as a broad $\sigma=0.2$\,keV
line centered at 6.69 keV and a broad $\sigma=0.1$\,keV line centered
at 2.38 keV, respectively, which reduced the residuals significantly.

The hard X-ray emission is modeled with a ``warm'' scattered powerlaw
reflector and a ``cold'' Compton-scattered continuum plus emission
lines. For the former component, we naively adopt a power law with the
same intrinsic slope as the obscured transmitted
component.\footnote{This warm mirror gas in \obj is possibly the same
  warm absorber gas seen in many Seyfert 1s (e.g,. K14), in which case
  the ionization level of the gas is not sufficiently high to be a
  perfect mirror. As noted in $\S$\ref{sec:intro}, this could imprint
  significant absorption edges/lines on the spectrum up to several keV
  \citep[e.g.,][]{Kaspi2002}, effectively adding spectral curvature,
  primarily below 2 keV, or flattening the slope of this component. We
  tested the possible effects of this modification on our results
  using an ionized absorber produced by {\sc xstar} for NGC\,3227
  \citep[see][for details]{Markowitz2009}. The primary effect was an
  increase in the normalization of the RRC component, with little
  change to the parameters of the components which dominate above
  2\,keV. Given this outcome, we chose a perfect mirror for
  simplicity. } For the latter component, we adopt either {\tt
  pexmon}, {\tt MYTorus}, or {\tt torus}, as described above, none of
which is particularly well-constrained by the \chandra{} HETG data
alone. All of the cold reflection models were smoothed with an
0.01~keV Gaussian to best-match the HETG neutral Fe K$\alpha$ line
width. We added neutral Ni K$\alpha$ (7.47\,keV) and/or K$\beta$
(8.23\,keV) lines when these were not modeled explicitly by the cold
reflection models; these lines are poorly constrained by the HETG
spectra, and thus were fixed relative to the full extraction region
values from \xmm{} assuming a nuclear to galaxy ratio of 2:1 as found
for neutral Fe K$\alpha$ (see $\S$\ref{sec:Fe_line}). Similar to the
transmitted component, we fixed the reflection component inclination
angle to $\theta_{\rm inc}=90$\degr{}, which is close to the nominal
viewing angle associated with \obj, and the high-energy exponential
cutoff rollover energy to $E_{\rm c}=$500\,keV for all
models. Finally, we included two neutral absorption ({\tt tbabs})
components, one of which was fixed at the Galactic column while the
other was fit as $N_{\rm
  H}=(1.5^{+0.2}_{-0.1})\times10^{21}$\,cm$^{-2}$ to constrain the
host column density in \obj.


We now proceed to fit the cold reflection with various prescriptions.
For all of the models below, we list the best-fit parameter values in
Table~\ref{tab:fits} (``Nucleus Only'') and show the resulting data-to-model
residuals in Figure~\ref{fig:hetg_nuc}. Fitting the {\tt pexmon} model
(model P in Table~\ref{tab:specmodels}) yielded a reduced
$\chi^{2}_{\nu}=1.60$ for $\nu=1471$ in the 0.5--9\,keV range. The
best-fit redshifts for the neutral and ionized lines were
0.00392$\pm0.0004$ and 0.00371$\pm0.00008$,
respectively, while the best-fit powerlaw index, Fe abundance, and
normalizations were 
$\Gamma=2.46^{\ ---}_{-0.24}$, 
$Z_{\rm Fe}=4.5^{+1.1}_{-0.6}$, 
$A_{\rm cold}=(8.9^{+5.0}_{-0.5})\times10^{-2}$ \pnorm{}, and 
$A_{\rm warm}=(2.8\pm0.2)\times10^{-4}$ \pnorm{},
respectively. The powerlaw slope is poorly constrained due to
parameter limitations of the {\tt pexmon} model.


We also fit the cold reflection with the {\tt MYTorus} model in two
distinct configurations (models M1 and M2 in
Table~\ref{tab:specmodels}). The first (M1) is a standard coupled
configuration, wherein the 
neutral hydrogen column densities $N_{\rm H}$,
intrinsic powerlaw slopes $\Gamma$,
inclination angles $\theta_{\rm inc}$,
and normalizations of the {\tt MYTZ} ($A_{\rm pow}$), {\tt MYTS}
($A_{\rm MYTS}$), and {\tt MYTL} ($A_{\rm MYTL}$) components are tied
and fit together self-consistently to model a uniform torus
geometry. The second (M2) is a decoupled configuration which employs
two Compton scatterers, one edge-on and one face-on, where the
corresponding normalizations for the different angles (e.g., $A_{\rm
  MYTS, 90}$ and $A_{\rm MYTS, 00}$) vary independently but the
continuum and line components of a given angle are fixed as in model
M1. This corresponds to a patchy torus whereby a portion of the
Compton-scattered photons which ``reflect'' off the facing side of
background clouds can bypass clouds which obscure photons along our
direct line of sight \citep[more details can be found in][and we refer
  interested readers particularly to their Figure 15]{Yaqoob2012}.

Fitting model M1 yielded a reduced $\chi^{2}_{\nu}=1.64$ for
$\nu=1472$ in the 0.5--9\,keV range. The best-fit redshifts for the
neutral and ionized lines were 0.00391$\pm0.0004$ and
0.00373$\pm0.00008$, respectively, while the best-fit powerlaw index,
scattering-to-line component (S/L) ratio,\footnote{I.e., the ratio of
  the $A_{\rm MYTS}$ to $A_{\rm MYTL}$ normalizations.}
$\Gamma=1.40^{+0.12}_{\ ---}$,
S/L ratio $=$ $0.42^{+0.12}_{-0.08}$,
$A_{\rm MYTS, cold}=7.4^{+2.9}_{-1.8}$ \pnorm{}, and
$A_{\rm warm}=(2.63\pm0.3)\times10^{-4}$ \pnorm{},
respectively. Meanwhile, model M2 yielded a reduced
$\chi^{2}_{\nu}=1.62$ for $\nu=1471$ in the 0.5--9\,keV range,
best-fit powerlaw index, scattering-to-line component ratio, and
normalizations were
$\Gamma=2.60^{\ ---}_{-0.19}$,
S/L ratio $=$ $0.67\pm0.09$,
$A_{\rm MYTS, 00}=0.19^{+0.01}_{-0.07}$ \pnorm{},
$A_{\rm MYTS, 90}=0.00^{+1.74}_{\ ---}$ \pnorm{}, and
$A_{\rm warm}=(4.1\pm0.2)\times10^{-4}$ \pnorm{},
respectively. As before, the powerlaw slope is not very
well-constrained over this particular energy range due to parameter
limitations of the {\tt MYTorus} model.


Finally, we fit the cold reflection with the {\tt torus} model (model
T in Table~\ref{tab:specmodels}). The best fit yielded a reduced
$\chi^{2}_{\nu}=1.65$ for $\nu=1472$ in the 0.5--9\,keV range.  The
best-fit redshifts for the neutral and ionized lines were
0.00363$^{+0.0004}_{-0.0003}$ and 0.00370$^{+0.00009}_{-0.00007}$,
respectively, while the best-fit powerlaw index, opening angle, and
normalizations were
$\Gamma=1.30^{+0.09}_{-0.05}$
$\theta_{\rm open}=67^{+12}_{-15}$ deg, 
$A_{\rm cold}=(2.6^{+0.8}_{-0.5})\times10^{-2}$ \pnorm, and 
$A_{\rm warm}=(3.0\pm0.2)\times10^{-4}$ \pnorm,
respectively. We note
that the relatively low $\Gamma$ value and small errors are largely
dictated by the Fe lines, since there is no way to change the Fe line
to continuum ratio through a metallicity parameter for this model.

As can be seen from Figure~\ref{fig:hetg_nuc} and
Table~\ref{tab:fits}, all of the models are able to fit the 0.5--9 keV
nucleus spectra equally well, with only very mild deviations in the
residuals between them. In all cases, the residuals are almost
exclusively due to low-level line emission (i.e., the strong ratio
outliers in the lower panels of Figure~\ref{fig:hetg_nuc}), most of
which is below 2 keV, that remains unaccounted for despite modeling
$\approx$90 emission lines. We found that these residuals bias the
relative normalization of the {\tt bremss} component downward by
$\approx$20\%, but do not appear to significantly affect the {\tt
  bremss} temperature nor normalizations of the higher energy
components (this holds for all of the cold reflection
models). Notably, there are wide variations in the powerlaw slopes
between models, which should be constrained better upon incorporating
the $>$10 keV data. If we limit the fit to the 2--9\,keV spectra and
fix the {\tt bremss} and {\tt tbabs} components, the reduced
$\chi^{2}_{\nu}$ values drop to $\approx$1 and the photon indices
become significantly harder ($\Gamma\approx$1.4--1.5) in all cases,
leading to decreased fractional contributions from the cold reflection
in the 2--10\,keV band. In the case of model M2, the 2--9 keV fit led
to a reversal in the dominant cold reflection component from 0\degr{}
to 90\degr{}. These large swings primarily demonstrate that the
spectral properties of the cold and warm reflection are poorly
constrained by the $<$10\,keV data alone, even when high
signal-to-noise and well-resolved emission lines can be fit.

\subsubsection{Diffuse Emission and Point Source Contamination From Host Galaxy}\label{sec:host}

Both extended and off-nuclear point source emission are evident in the
\chandra{} images, particularly along the direction of the AGN radio
jet and counter jet (see Figure~\ref{fig:images}). We modeled this
emission in the \chandra{} ACIS-S data with several components to
reproduce the main features in the galaxy, noting in particular that
there are several key spectral signatures present in the nuclear
spectra which are also prevalent in the host spectrum.

\begin{figure}[t]
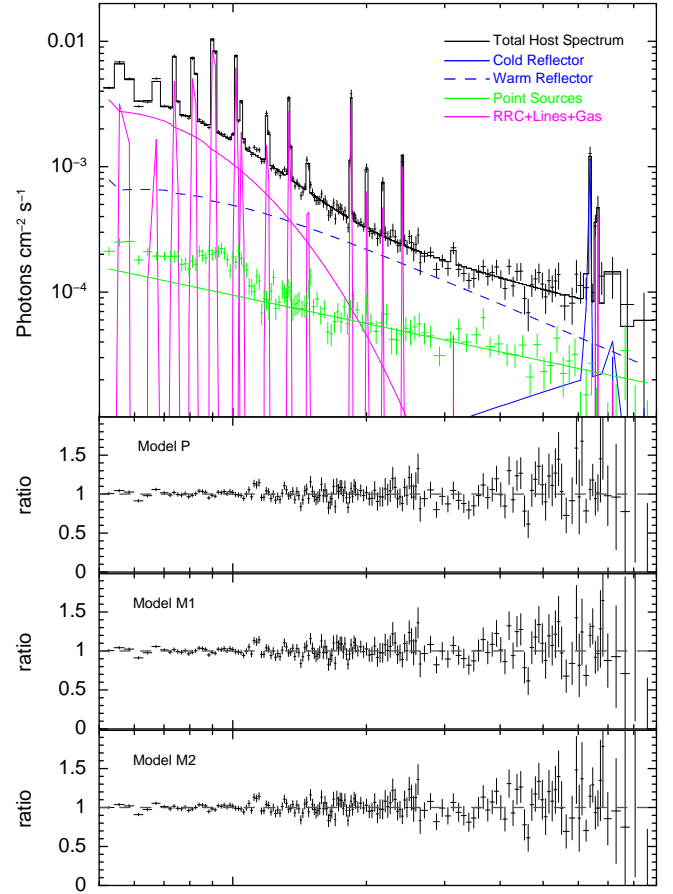

\vspace{0.5cm}
\hglue-0.5cm{\includegraphics[angle=270, width=9.0cm]{Model_P_host_ACISS.ps}}
\vglue-0.09cm{\hglue-0.75cm{\includegraphics[angle=270, width=9.25cm]{Model_P_host_ACISS2.ps}}}
\vglue-0.13cm{\hglue-0.75cm{\includegraphics[angle=270, width=9.25cm]{Model_M1_host_ACISS.ps}}}
\vglue-0.13cm{\hglue-0.75cm{\includegraphics[angle=270, width=9.25cm]{Model_M2_host_ACISS.ps}}}
\vspace{0.0cm} \figcaption[]{ The total ACIS-S ``host'' unfolded
  spectrum (black) of all extranuclear emission extracted from a
  2--75\arcsec{} annulus centered on the nucleus of \obj. The host
  spectrum is fitted with four models comprised of partially absorbed
  cold and warm reflectors (blue solid and dashed lines,
  respectively), as well as RRC and the most prominent line components
  as identified by K14 (magenta solid lines), and the composite
  contribution from extranuclear point sources (green solid line). The
  top panel shows the fit to model P while the rest of the panels show
  the data-to-model ratios for models P, M1, and M2 respectively. The
  top panel also shows the composite ACIS-S extranuclear point-source
  spectrum (green), which is modeled above 1.5~keV as a $\Gamma_{\rm
    ULX}=1.2$ power law with a $E_{\rm c}=7$\,keV exponential cutoff
  rollover.
\label{fig:host}}
\vspace{0.5cm} 
\end{figure}

First, we include in the host galaxy model an absorbed power law with
slope $\Gamma_{\rm pnt}$ to account for the combined emission from
extranuclear point sources, which we constrain separately below.  A
composite \chandra{} ACIS-S spectrum of all of the point sources
together is shown in Figure~\ref{fig:host} (green data and
model). There are some notable bumps in the soft portion of the
spectrum, which could either be intrinsic or more likely are produced
by poor background subtraction due to an inhomogenous extended
emission component. As such, we fitted this spectrum only above
1.5\,keV with a single cutoff powerlaw model.  Unfortunately, the
limited 0.5--9\,keV energy range is not sufficient to unambiguously
determine the average spectrum slope, high-energy cutoff, and
normalization of the host galaxy point-source population. Following
\citet{Swartz2004} and \citet{Walton2011}, we assume that \obj hosts
an ultraluminous X-ray source (ULX) population and that emission
characteristic of this population likely dominates the point source
emission. Recent evidence from \nustar{} \citep[e.g.,
][]{Bachetti2013, Rana2014, Walton2013, Walton2014} suggests that ULXs
exhibit relatively hard spectra with spectral turnovers between
6--8\,keV, and thus we adopt fixed values of $\Gamma=1.2$ and $E_{\rm
  c}=7$\,keV to represent the composite ULX-like spectrum. With these
values, the normalization of the power law is $8.9\times10^{-5}$
\pnorm{}. This component makes only a relatively small contribution to
the overall host contamination in the 1.5--9.0 keV ($\approx$25\%)
range and quickly becomes negligible above 15\,keV. We fixed the
normalization of this fit and added this fixed off-nuclear
point-source component to the overall host model.

At soft energies, we still see signs of extended RRC and line
emission, which we again model as a $kT_{\rm
  bremss}=0.31^{+0.01}_{-0.01}$\,keV bremsstrahlung component ({\tt
  bremss}, $A_{\rm bremss}=0.0168^{+0.0004}_{-0.0003}$ cm$^{-2}$) plus
a subset of the 20 strongest emission lines found in the nuclear
spectra; at the spectral resolution of ACIS-S, these 20 lines were
sufficient to model nearly all of the spectral deviations from a
smooth continuum. There may also be a contribution from hot gas
associated with star formation, but since our main focus is to derive
an empirical model to describe the soft emission, we simply absorb
this into the normalization for the bremsstrahlung plus line emission
model. The character of the ionized lines differs from those found in
the nucleus spectrum, in the sense that lower ionization line species
such as S, Si, Mg are stronger in the host spectra relative to the
ionized Fe lines, as might be expected for a UV/X-ray radiation field
which is radiating from the central SMBH.

At hard energies, we additionally see traces of warm and cold AGN
reflection as extended emission, which we model as a scattered power
law and Compton-scattered continuum plus neutral lines,
respectively. We continue to model the latter with either the {\tt
  pexmon} or {\tt MYTS}$+${\tt MYTL}; we do not fit the {\tt torus}
model, since one cannot explicitly separate out the transmitted
component.\footnote{The {\tt torus} transmitted component could be
  made negligible by increasing the column density to
  $10^{26}$\,cm$^{-2}$, but this would mean we would have to model all
  clouds as extremely Compton-thick, which is a major limitation.}  As
before, we will assume that the warm and cold reflection components
result from the scattering of the same direct transmitted power law
({\tt cutoffpl}) with slope $\Gamma_{\rm nuc}$ and exponential cutoff
rollover energy $E_{\rm c,nuc}$, which is absorbed along the line of
sight by a Compton-thick absorber (e.g., an edge-on torus). As before, we
fixed the quantities $E_{\rm c,nuc}=500$\,keV, $\theta_{\rm
  inc}=90$\degr{} and $N_{\rm H}=10^{25}$ cm$^{-2}$, since these are
poorly constrained by the $<$10\,keV data alone.

Finally, we note that the absorption toward the counter-jet region is
significantly stronger than that toward the jet region, so we
initially fit all the components to the jet and counter-jet regions,
allowing only for the $N_{\rm H}$ of the cold absorber to vary between
them. This fit produced $N_{\rm H}=3.1\times10^{20}$\,cm$^{-2}$ toward
the jet, consistent with the Galactic column, and $N_{\rm
  H}=2.4\times10^{21}$\,cm$^{-2}$ toward the counter jet. As such, the
2--75\arcsec{} host region was modeled through a layer of cold
Galactic absorption ({\tt tbabs}) and a cold partial coverer ({\tt
  pcfabs}) with $N_{\rm H}=2.4\times10^{21}$\,cm$^{-2}$ and covering
fraction of 50\%. For all of the models, we list the best-fit
parameter values in Table~\ref{tab:fits} (``Host Only'') and show the
resulting data-to-model residuals in Figure~\ref{fig:host}.


Fitting the {\tt pexmon} (P) version of our host model yielded a
reduced $\chi^{2}_{\nu}=1.42$ for $\nu=163$ in the 0.5--9\,keV
range. Given the quality and spectral resolution of the ACIS-S
spectrum, we fixed the redshift at 0.00379. The best-fit powerlaw
index, Fe abundance, and normalizations were
$\Gamma=2.49^{\ ---}_{-0.25}$, 
$Z_{\rm Fe}=43^{\ ---}_{-19}$, 
$A_{\rm cold}=(2.5^{+0.4}_{-0.5})\times10^{-2}$ \pnorm, and 
$A_{\rm warm}=(6.7\pm0.2)\times10^{-4}$ \pnorm, 
respectively. It is worth noting here that the abundance value, albeit
poorly constrained, is exceptionally high and probably highlights a
critical breakdown of the model in this regime rather than an extreme
intrinsic value.
We also fit the host spectrum with the {\tt MYTorus} (M1 and M2)
versions of our host model. Fitting model M1 produced a reduced
$\chi^{2}_{\nu}=1.44$ for $\nu=163$ in the 0.5--9\,keV range. 
The best-fit powerlaw index, scattering-to-line
component ratio, and normalizations were
$\Gamma=2.55^{\ ---}_{-0.06}$,
S/L ratio of $2.46^{+3.49}_{-1.01}$,
$A_{\rm MYTS, cold}=1.2^{+0.8}_{-0.7}$ \pnorm, and
$A_{\rm warm}=(6.8\pm0.2)\times10^{-4}$ \pnorm, 
respectively. Meanwhile, model M2 yielded a reduced
$\chi^{2}_{\nu}=1.46$ for $\nu=162$ in the 0.5--9\,keV range,
best-fit powerlaw index, scattering-to-line component ratio, and
normalizations were
$\Gamma=2.56^{\ ---}_{-0.05}$,
S/L ratio of $2.25^{+2.65}_{-0.90}$,
$A_{\rm MYTS, 00}=(1.7^{+1.1}_{\ ---})\times10^{-2}$ \pnorm,
$A_{\rm MYTS, 90}=0.00^{+1.58}_{\ ---}$ \pnorm, and 
$A_{\rm warm}=(6.7\pm0.2)\times10^{-4}$ \pnorm,
respectively. We note that the reflection component from the host
emission should be comprised almost exclusively of inclination 0\degr{}
(``far-side, face-on'') reflection spectra whose line-of-sight does not
intercept any torus material \citep[see further discussion
  in][]{Yaqoob2012}; thus we can effectively neglect the 90\degr{}
component altogether.

Similar to the nucleus fits, the powerlaw slopes for models P, M1, and
M2 were not well-constrained due to parameter limitations of the
various models and data bandpass limitations. The bulk of the
residuals arise from unaccounted-for line emission below 2 keV. As
seen in Table~\ref{tab:fits}, when we fit the models to the $>$2 keV
spectrum and fix the bremsstrahlung component, the reduced
$\chi^{2}_{\nu}$ values drop considerably for all models.

\subsubsection{Empirical Constraints on Extended Fe Line Emission}\label{sec:Fe_line}

An alternative, more empirical approach can be made to understand the
contribution from extended cold and warm reflection. For this, we
simply measure the line fluxes from the two strongest tracers, the
fluorescent Fe K$\alpha$ line and the ionized Fe He-like line,
respectively.  For simplicity, we use the M04a model (although we
replace {\tt pexmon} by {\tt pexrav} in order to remove emission lines
from the model) to estimate the continuum in both the \chandra{} HETG
nuclear and ACIS-S host spectra, and then model the remaining lines
with Gaussians as before in $\S$\ref{sec:previous}. The line fluxes
from the nuclear and host spectra are shown in Table~\ref{tab:lines}
alongside the total line fluxes measured from the pn
spectra. Reassuringly, the sum of the nuclear plus host are consistent
with the total line fluxes, at least when we factor in statistical
errors and cross-calibration differences.

After we account for contributions from the extended wings of the PSF
using simulations from the {\sc
  marx}\footnote{http://space.mit.edu/cxc/marx/} ray-trace simulator
\citep[v4.5;][]{Wise1997}, we find that the extended Fe K$\alpha$
emission beyond 2\arcsec{} ($>$140 pc) comprises 28$^{+8}_{-8}$\% of
the total. If the torus size is of order $\approx$4--10 pc, then we
should probably consider the extended fraction above to be a lower
limit to the cold reflection contribution from extended (i.e,
non-torus) clouds, since there are likely to be contributions from
similar material at 10--140 pc. Making a similar calculation for the
ionized Fe He-like line, we find an extended fraction of
24$^{+18}_{-20}$\%.



\subsubsection{Combined Fit}\label{sec:combined}

\begin{figure*}[t]
\vspace{-0.0in}
\hglue-0.5cm{\includegraphics[angle=270, width=18cm]{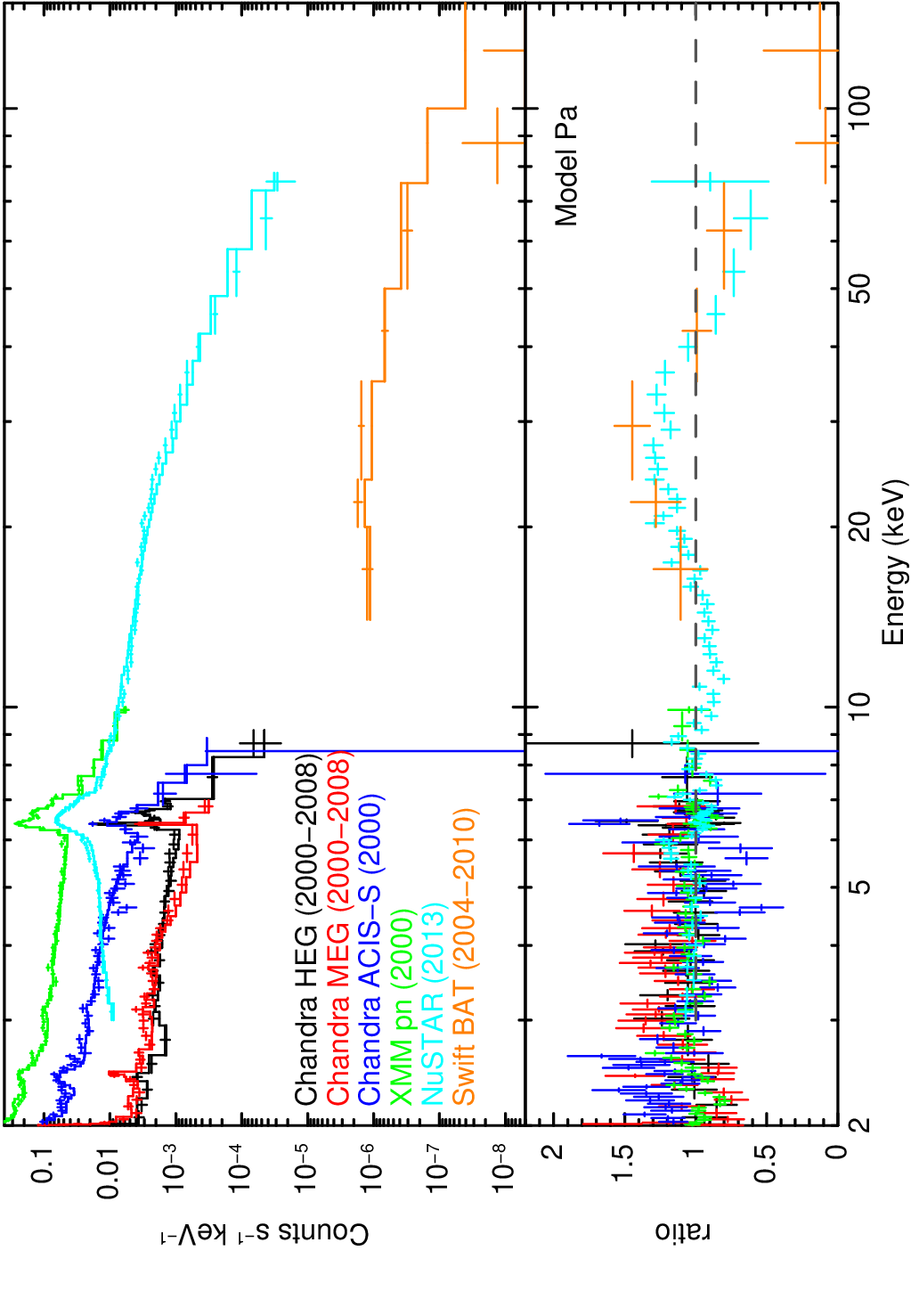}}
\vspace{0.0cm} \figcaption[]{The top panel shows the final selection
  of X-ray spectra for \obj that we fitted from \nustar{} FPMA/FPMB
  (cyan), \xmm{} pn (green), \chandra{} HEG/MEG (black/rad),
  \chandra{} ACIS-S (blue), and \swift{} BAT (orange), all modeled
  with the best-fit parameters from model Pa, while the bottom panel
  shows the data-to-model ratios for each spectrum. As in
  Figure~\ref{fig:M04_fix_allhard}, the overall consistency between
  the various datasets is good, once known normalization offsets are
  accounted for. In particular, the sum of the \chandra{} HEG/MEG
  (``Nucleus only'') and ACIS-S (``Host only'') models provides an
  excellent fit to the other (``Total'') datasets where they overlap
  in energy. The only discrepancy between datasets appears to be a
  broad deficiency between 5.5--6.1\,keV for the \nustar{} data. Model
  Pa is similar in shape to the M04a model, which still provides a
  poor fit to the data near the rise and peak of the Compton
  reflection hump. The HETG spectra are rebinned for presentation
  purposes.
\label{fig:Pa_data}}
\vspace{0.5cm} 
\end{figure*}

\begin{figure}[t]
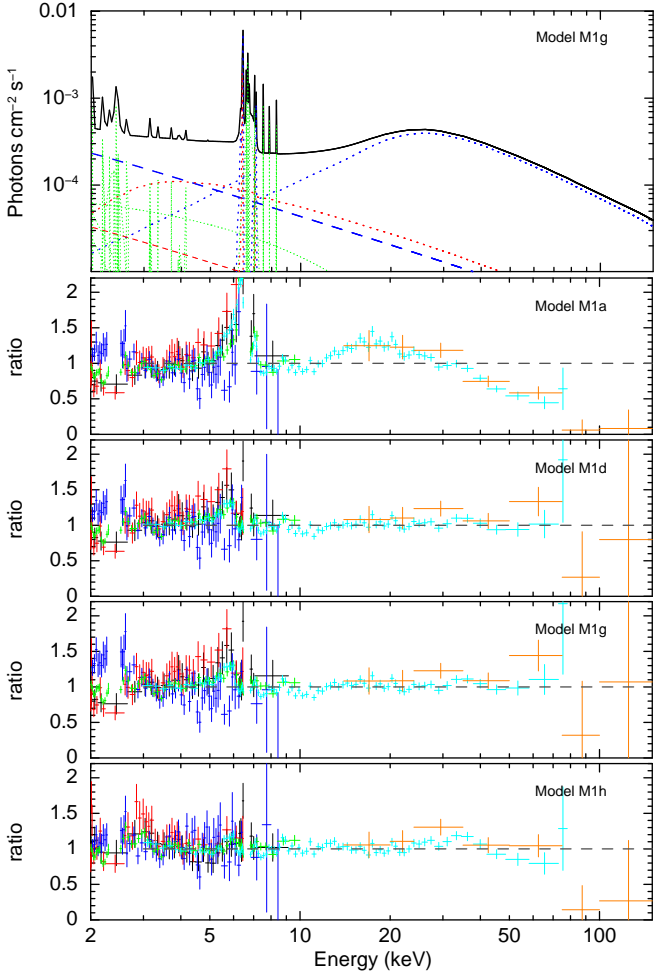

\vspace{-0.0in}
\vglue-0.09cm{\hglue-0.77cm{\includegraphics[angle=270, width=9.08cm]{Model_M1g_total_model.ps}}}
\vglue-0.00cm{\hglue-0.7cm{\includegraphics[angle=270, width=9cm]{Model_M1a_total_ratio.ps}}}
\vglue-0.08cm{\hglue-0.7cm{\includegraphics[angle=270, width=9cm]{Model_M1d_total_ratio.ps}}}
\vglue-0.08cm{\hglue-0.7cm{\includegraphics[angle=270, width=9cm]{Model_M1g_total_ratio.ps}}}
\vglue-0.08cm{\hglue-0.7cm{\includegraphics[angle=270, width=9cm]{Model_M1h_total_ratio.ps}}}
\vspace{0.0cm} \figcaption[]{ {\it Top panel}: Model M1g shown for the full dataset. Solid lines denote the overall spectrum.
  Blue lines represent the nuclear warm (dashed) and cold (dotted)
  reflection components.
%
  The red lines represent the host warm (dashed) and cold (dotted)
  reflection components,
  The RRC and line emission components for both the nuclear and host
  models are shown as dotted green lines.
  {\it Bottom panels}: Data-to-model ratios for several M1 models,
  with the same color-coding as Figure~\ref{fig:Pa_data}. Many of the 
  models we fit exhibited poor fits to the data either above or below
  10\,keV. The HETG spectra are rebinned for presentation purposes.
\label{fig:M1_fullmods}}
\vspace{0.5cm} 
\end{figure}

\begin{figure}[t]
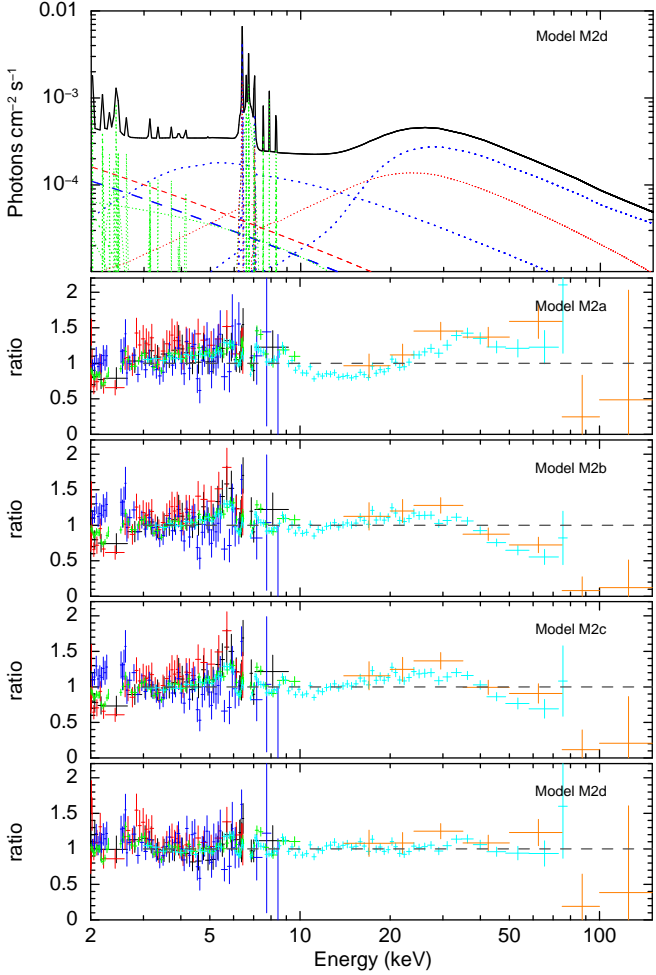

\vspace{-0.0in}
\vglue-0.09cm{\hglue-0.77cm{\includegraphics[angle=270, width=9.08cm]{Model_M2d_total_model.ps}}}
\vglue-0.00cm{\hglue-0.7cm{\includegraphics[angle=270, width=9cm]{Model_M2a_total_ratio.ps}}}
\vglue-0.08cm{\hglue-0.7cm{\includegraphics[angle=270, width=9cm]{Model_M2b_total_ratio.ps}}}
\vglue-0.08cm{\hglue-0.7cm{\includegraphics[angle=270, width=9cm]{Model_M2c_total_ratio.ps}}}
\vglue-0.08cm{\hglue-0.7cm{\includegraphics[angle=270, width=9cm]{Model_M2d_total_ratio.ps}}}
\vspace{0.0cm} \figcaption[]{ Same as Figure~\ref{fig:M1_fullmods}, but for M2 models. The model M2d provides the best overall fit to the spectra among all of the models. \label{fig:M2_fullmods}}
\vspace{0.5cm} 
\end{figure}

\begin{figure}[t]
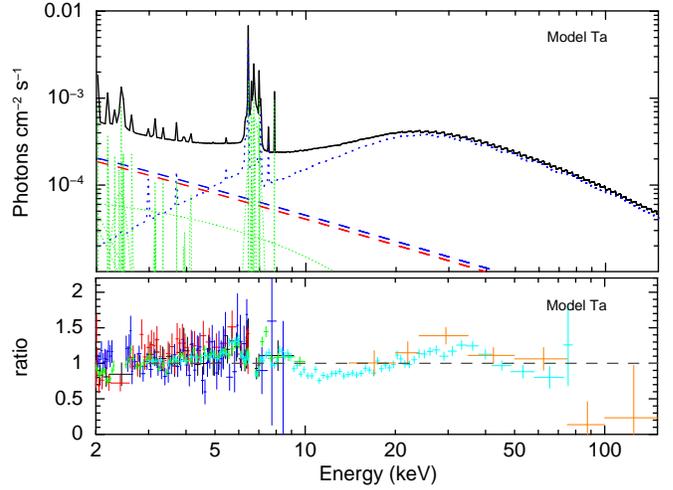

\vspace{-0.0in}
\vglue-0.09cm{\hglue-0.77cm{\includegraphics[angle=270, width=9.08cm]{Model_Ta_total_model.ps}}}
\vglue-0.00cm{\hglue-0.7cm{\includegraphics[angle=270, width=9cm]{Model_Ta_total_ratio.ps}}}
\vspace{0.0cm} \figcaption[]{ Same as Figure~\ref{fig:M1_fullmods}, but for T models. \label{fig:T_fullmods}}
\vspace{0.5cm} 
\end{figure}

We now combine the models of the nucleus and host galaxy from the
\chandra{} spectra to fit the total spectra from \nustar{}, \xmm{},
and \swift{} BAT.  As highlighted previously, the emission below
$\approx$2\,keV is dominated by the numerous line and bremsstrahlung
components, and thus does not provide much constraint on the
properties of the reflectors. At the same time it contributes
substantially to $\chi^{2}$, so for the remainder of the modeling we
only consider the data above 2 keV. All of the spectral components
that are well-constrained by the previous nuclear and host spectral
fitting, such as the extranuclear point-source, RRC and line emission,
are fixed, as we are primarily concerned with constraining the
relative contributions from the warm and cold reflection, as well as
any potential direct AGN continuum. For modeling simplicity, we also
chose to ignore the regions between 2.3--2.5\,keV and 6.5--6.8\,keV,
which correspond to regions of ionized Si and Fe line emission,
respectively; these regions always have considerable residuals which
are not modeled by the continuum reflection components but bias the
component normalizations during the fitting process. We assume below
that all of the components share a single intrinsic powerlaw slope and
that any transmitted component, if present, must arise only from the
nuclear portion of the spectrum. For selected relevant models below,
we list the best-fit parameter values in Table~\ref{tab:fits}
(``Total'') and/or plot their residuals in
Figures~\ref{fig:Pa_data}--\ref{fig:T_fullmods}.

\paragraph{Model P}\label{sec:modelP}

We begin by fitting model P to the combined 2--195\,keV spectra of
\obj.  We fit $\Gamma$ and $Z_{\rm Fe}$, as well as the normalizations
$A_{\rm cold,nuc}$, $A_{\rm cold,host}$, $A_{\rm warm,nuc}$ and
$A_{\rm warm,host}$ as free parameters, while we fix $\theta_{\rm
  inc}=85$\degr{}, $E_{\rm c}=500$~keV and $N_{\rm
  H}=10^{25}$~cm$^{-2}$.  This model, hereafter ``Pa'', yielded a poor
fit, with a reduced $\chi^{2}_{\nu}=1.34$ for $\nu=1666$. The Pa model
residuals, which are shown in Figure~\ref{fig:Pa_data}, highlight a
general problem with fitting the spectral shape above 8\,keV that we
encountered with many of the adopted models, namely that the models
either fit the spatially resolved $<10$\,keV data well but present
clear $>10$\,keV residuals, or vice versa. Allowing the cutoff energy
to vary failed to yield any improvement in $\chi^{2}_{\nu}$, with a
best-fit value of $E_{\rm c}=500^{\ ---}_{-176}$\,keV, hereafter model
``Pb''. Alternatively, allowing the inclination angle to also vary to
$\theta_{\rm inc}=24^{+7}_{-5}$ deg, hereafter model ``Pc'',
significantly improved the fit, with a new reduced
$\chi^{2}_{\nu}=1.28$. We note that this inclination angle suggests a
face-on configuration, perhaps indicative of scattering off of the
back wall of a fiducial torus, while the best-fit photon index
($\Gamma=1.65\pm0.02$) is somewhat lower than one would expect for
such a high accretion rate source like \obj \citep[e.g.,
  $\Gamma=2.5$;][]{Fanali2013}. Critically, although the high-energy
residuals have improved, significant deviations of the form shown in
Figure~\ref{fig:Pa_data} from the observed continuum shape still
remain. Again, varying the cutoff energy to $E_{\rm
  c}=387^{\ ---}_{-176}$\,keV fails to yield any substantial
improvement in $\chi^{2}_{\nu}$.


\paragraph{Model M1}\label{sec:modelM1}

We now turn to the cold reflection as modeled by {\tt MYTorus}. As
before, we initially adopt a ``standard'' fully coupled, uniform torus
geometry, hereafter ``M1a''. While there is no physical reason for the
nuclear and extended components to be the same, we begin with such a
scenario because it represents how previous studies would model the
entire \xmm{} or \nustar{} spectrum. For the M1a model, we fit
$\Gamma=1.40^{0.09}_{\ ---}$ and the component normalizations, and fix
the other parameters to $N_{\rm H}=10^{25}$\,cm$^{-2}$, $\theta_{\rm
  inc}=90$\degr{}, $E_{\rm cut}=500$\,keV, and the S/L ratio to 1.
Aside from allowing the reflection component normalazations to vary,
the properties of the nucleus and host reflectors were tied
together. The resulting fit was poor, with $\chi^{2}_{\nu}=3.78$ for
$\nu=1666$, and large residuals around both the neutral Fe K$\alpha$
line and to a lesser extent the Compton hump. Moreover, the powerlaw
slope is quite flat. From the residuals, it is clear that a S/L ratio
of 1 is insufficient, and allowing the S/L ratio to vary to
26.7$^{+14.2}_{-1.0}$, hereafter ``M1b'', substantially improved the
fit with $\chi^{2}_{\nu}=1.78$. Such a S/L ratio is unreasonbly high,
however, and implies that the adopted values for some of the fixed
parameters are likely wrong. Varying $E_{\rm cut}$ to
$55^{+4}_{-5}$\,keV (``M1c'') lowered the S/L ratio to
15.0$^{+1.2}_{-0.9}$ and resulted in $\chi^{2}_{\nu}=1.61$.  Finally,
further varying the inclination angle and column density improves the
fit to $\chi^{2}_{\nu}=1.31$, with $\Gamma=1.40^{+0.12}_{\ ---}$,
$N_{\rm H}=(9.4^{\ ---}_{-3.3})\times10^{24}$\,cm$^{-2}$, $\theta_{\rm
  inc}=78^{+3}_{-4}$, $E_{\rm cut}=41^{+5}_{-4}$\,keV, and an
S/L$_{\rm nuc+host}$ ratio of 3.8$^{+0.5}_{-0.8}$ (``M1d''). This last
model fits the $>$10\,keV continuum significantly better, but at the
expense of producing residuals in the $<$10\,keV continuum (see
Figure~\ref{fig:M1_fullmods}) while retaining a flat powerlaw
slope. Ultimately, we conclude that none of the coupled {\tt MYTorus}
models provides a reasonable fit to the continuum shape. It is
important to point out that if we had only modeled either the
$<$10\,keV spectra or the total aperture spectra, we would have
arrived at a satisfactory $\chi^{2}_{\nu}$.

As an alternative to the fully coupled models, we tried fitting
separate {\tt MYTS}+{\tt MYTL} parameters for the nucleus and the host
spectra, as might be expected for the combination of a thick torus and
more tenuously distributed larger scale molecular clouds, which has
been found from mid-IR constraints on \obj. We began by fitting a
single photon index $\Gamma=1.80^{+0.05}_{-0.07}$, the various
component normalizations, and independent column densities $N_{\rm
  H,nuc}=(9.8^{\ ---}_{-0.2})\times10^{24}$ cm$^{-2}$ and $N_{\rm
  H,host}=(2.4^{+0.1}_{-0.2})\times10^{23}$ cm$^{-2}$ and S/L ratios
12.2$^{+1.8}_{-1.9}$ and 0.5$^{+0.3}_{-0.2}$ for the nucleus and host
components, respectively, while fixing $\theta_{\rm inc}=90$\degr{}
and $E_{\rm cut}=500$\,keV (``M1e''). This fit produced a reduced
$\chi^{2}_{\nu}=1.54$ for $\nu=1663$. Allowing $E_{\rm
  cut}=33^{+5}_{-3}$\,keV improved the fit to $\chi^{2}_{\nu}=1.30$,
with modest changes to the other free parameters such that the
$\Gamma$ remained pinned at its minimum while $N_{\rm
  H,nuc}=(5.3^{+0.4}_{-0.5})\times10^{24}$ cm$^{-2}$, $N_{\rm
  H,host}=(0.09\pm0.03)\times10^{24}$ cm$^{-2}$, S/L$_{\rm
  nuc}=10.8^{+1.3}_{-0.8}$ and $_{\rm host}=1.0^{+0.4}_{-0.2}$
(``M1f''). Finally, allowing the inclination angles to vary (``M1g'')
only marginally improves the fit to $\chi^{2}_{\nu}=1.28$, with free
parameters $\Gamma=1.40^{+0.34}_{\ ---}$, $E_{\rm
  cut}=34^{+58}_{-4}$\,keV, $N_{\rm
  H,nuc}=(8.0^{\ ---}_{-1.6})\times10^{24}$ cm$^{-2}$, $N_{\rm
  H,host}=(1.3^{+1.5}_{-0.9})\times10^{24}$ cm$^{-2}$, S/L$_{\rm nuc}=
3.5^{+0.8}_{-0.5}$ and S/L$_{\rm host}=1.5\pm0.3$.

We note that freeing the column density and normalization toward the
transmitted component (``M1h'') to $N_{\rm
  H,trans}=(6.0^{+1.3}_{-0.8})\times10^{24}$ cm$^{-2}$ results in a
reduced $\chi^{2}_{\nu}=1.13$, with best-fit values of
$\Gamma=2.20^{+0.07}_{-0.12}$, $E_{\rm cut}=72^{+75}_{-21}$\,keV,
$N_{\rm H,nuc}=(2.6^{+0.5}_{-0.5})\times10^{23}$ cm$^{-2}$ and $N_{\rm
  H,host}=10^{25}$ cm$^{-2}$ (unconstrained), S/L$_{\rm
  nuc}= 1.0^{+0.2}_{-0.3}$ and S/L$_{\rm host}=2.0^{+0.6}_{-0.6}$, and
inclination angles of 0.7$^{+4.5}_{\ ---}$ deg and 1.9$^{+10.5}_{\ ---}$
deg for the nucleus and host components, respectively. This model is
the best version of the ``standard'' {\tt MYTorus} configuration and
crudely models the key continuum and line features, but ultimately
predicts that \obj should be dominated by the tranmitted component
above 20 keV. The normalizations of the various continuum components
are $A_{\rm trans}=2.6^{+1.3}_{-1.3}$ \pnorm, $A_{\rm
  warm,nuc}=(3.0^{+1.3}_{-1.3})\times10^{-4}$ \pnorm, $A_{\rm
  cold,nuc}=(4.0^{+0.7}_{-0.5})\times10^{-2}$ \pnorm, $A_{\rm
  warm,nuc}=(3.9^{+1.5}_{-1.4})\times10^{-4}$ \pnorm, $A_{\rm
  cold,nuc}=(9.4^{+4.2}_{-3.7})\times10^{-3}$ \pnorm, respectively,
impling a covering fractions of $\sim$0.008 and $\sim$0.002 for the
nucleus and host cold reflection components. Such low covering
fractions run contrary to the variability constraints presented in
$\S$\ref{sec:data_nustar} and $\S$\ref{sec:data_swift}. As such, the
good fit appears to be a consequence of allowing freedom for several
spectral components to fit small portions of the overall spectrum, and
is presumably degenerate in this sense.

We conclude that the ``standard'' configuration of {\tt MYTorus} has
considerable difficulty reproducing the main spectral and temporal
X-ray characteristics of \obj.

\paragraph{Model M2}\label{sec:modelM2}

We now turn to the second {\tt MYTorus} configuration, which employs
two {\tt MYTorus} Compton scatterers fixed at 0\degr{} and 90\degr{},
representing a potential clumpy torus-like distribution. Following the
discussion in $\S$\ref{sec:host}, we only invoke the 0\degr{}
component to fit the host spectrum. We began by fitting a basic form
of this model, hereafter ``M2a'', with varying
$\Gamma=2.29^{+0.04}_{-0.02}$ and component normalizations with the
remaining parameters fixed to $N_{\rm H}=10^{25}$\,cm$^{-2}$,
S/L$_{\rm nuc+host}=1.0$, and $E_{\rm cut}=500$\,keV for all
scattering components. The best fit returns a $\chi^{2}_{\nu}=1.84$
for $\nu=1666$, which is a significant improvement over model M1a.
However, the continuum is still not well-fit and the best-fit $A_{\rm
  nuc, MYTS, 90}$ normalization is consistent with zero ($<$1\% of
cold reflector flux).
Fitting the S/L$_{\rm nuc+host}$ ratio to 4.3$^{+0.4}_{-0.3}$
(``M2b'') reduces $\chi^{2}_{\nu}=1.51$, and yields 
$\Gamma=1.49\pm0.04$ plus moderate variations in the component
normalizations.  M2b offers a significant improvement over model
M1b.
Additionally varying $E_{\rm cut}=146^{+76}_{-50}$\,keV (``M2c''),
provides only very marginal improvement ($\chi^{2}_{\nu}=1.48$) and
leaves the parameters largely unmodified.
Finally, varying the three column densities (``M2d'') improves the fit
to $\chi^{2}_{\nu}=1.14$, with $\Gamma=2.10^{+0.06}_{-0.07}$, a
S/L$_{\rm nuc+host}$ ratio of 1.0$\pm0.1$, $E_{\rm cut}=128^{+115}_{-44}$\,keV
$N_{\rm H,nuc,90}=(10.0^{\ ---}_{-4.4})\times10^{24}$\,cm$^{-2}$, 
$N_{\rm H,nuc, 0}=(1.5\pm0.1)\times10^{23}$\,cm$^{-2}$, 
$N_{\rm H,host,0}=(5.0^{4.5}_{-1.9})\times10^{24}$\,cm$^{-2}$, 
and normalizations of
$A_{\rm warm,nuc}=(2.5^{+0.3}_{-0.4})\times10^{-4}$ \pnorm, 
$A_{\rm cold,nuc,90}=(3.0\pm0.5)\times10^{-1}$ \pnorm, 
$A_{\rm cold,nuc, 0}=(3.6^{+0.3}_{-0.2})\times10^{-2}$ \pnorm, 
$A_{\rm warm,host}=(3.4^{+0.3}_{0.4})\times10^{-4}$ \pnorm, 
$A_{\rm cold,host,0}=(1.0\pm0.2)\times10^{-2}$ \pnorm.
Freezing the high-energy cutoff at $E_{\rm cut}=500$\,keV (``M2e'') leaves the above
parameters virtually unchanged and $\chi^{2}_{\nu}=1.16$.

As can be seen in Figure~\ref{fig:M2_fullmods}, the data-to-model
ratio residuals are now fairly flat out to $\approx$80\,keV. The
primary difference between model M2d (or M2e) and all of the others
lies in how the nuclear $\theta_{\rm inc}=0$\degr{} cold reflector
component, due to its significantly lower $N_{\rm H}$, is able to fill
in the spectral gap around 4--8\,keV between the ``normal'' cold and
warm reflectors. One important aspect of this model which deserves
highlighting is the fact that while the higher $N_{\rm H}$ component
provides the bulk of the flux to the Compton hump, it does not
contribute much to the Fe fluoresence line emission. Instead, the
lower $N_{\rm H}$ component produces the bulk of the Fe fluoresence
line emission and dominates the continuum peaking around
5--10\,keV. Thus the two key features of Compton reflection, namely
the hump and Fe line, need not arise from a single absorber and in
fact likely arise from different obscuring clouds.  Assuming a single
absorber will likely lead to misinterpretations.


\paragraph{Model T}\label{sec:modelT}

Finally, we fit the cold reflection with the {\tt torus} model.  As
noted in $\S$\ref{sec:new}, this model is not suitable for fitting the
host spectrum, so we instead modeled the host spectrum identically to
the M2 case using {\tt MYTS}+{\tt MYTL} components with an inclination
angle of $\theta_{\rm inc}=0$\degr{}. Varying 
$\Gamma=1.96^{+0.05}_{-0.04}$, 
$\theta_{\rm open}=64^{+3}_{-2}$ deg, 
and component normalizations, with fixed values of 
$N_{\rm H}=10^{25}$\,cm$^{-2}$, $\theta_{\rm inc}=87$\degr{},
$E_{\rm cut}=500$\,keV, and a S/L$_{\rm host}$ ratio of 1.0, yielded a
reduced $\chi^{2}_{\nu}=1.61$ for $\nu=1666$ (``Ta''). This provides a
relatively poor fit, with residuals near the Fe lines and $>$10 keV
continuum (Figure~\ref{fig:T_fullmods}). Freeing the {\tt torus}
inclination angle to $\theta_{\rm T, inc}=87^{\ ---}_{-16}$ (``Tb'')
does not improve the fit. Further varying the nuclear and host column
densities to $N_{\rm
  H,nuc}=(6.9^{+0.6}_{-0.8})\times10^{24}$\,cm$^{-2}$, $N_{\rm H,host,
  0}=(10.0^{\ ---}_{-6.6})\times10^{24}$\,cm$^{-2}$ (``Tc'') leads to
a modest improvement $\chi^{2}_{\nu}=1.57$, with
$\Gamma=2.13^{+0.04}_{-0.06}$, $\theta_{\rm open}=69^{+4}_{-3}$ deg,
and $\theta_{\rm incl}=87^{\ ---}_{-12}$ deg.
As with other models, there are significant residuals as the model
fails to fit the continuum shape well. In all cases, the host cold
reflection normalization is consistent with zero.  It seems that the
{\tt torus} model does not provide enough flexibility to model the
transmission and scattered components separately and again we conclude
that the {\tt torus} model has considerable difficulty reproducing the
main spectral X-ray characteristics of \obj.

\subsubsection{Model Summary}

We tested a variety of cold reflection models earlier in this
section. As has been traditionally done in the past, we modeled \obj
with a single monolithic cold reflector using {\tt pexmon} (models
Pa--Pc), {\tt MYTorus} (models M1a--M1d), and {\tt torus} (models
Ta-Tc). Alternatively, we also modeled \obj with multiple reflectors
using two or three {\tt MYTorus} components to fit the two spatially
distinct nuclear and host regions (models M1e--M1h) and additional
complexity in the nuclear spectrum (models M2a-M2d). We found that
many models are able to fit either the spatially resolved $<10$\,keV
spectra or the total aperture spectra well, but generally not both.

The two models which do manage to fit all of the spectra well are M1h
and M2d. In both cases, a cold reflection component with \hbox{$N_{\rm
    H}\sim10^{23}$~cm$^{-2}$} peaking at 5--10\,keV is required to
fill in a critical gap in the model where the declining warm reflector
and the increasing cold reflector meet. Model M1h is rejected,
however, because it requires a strong transmitted component, which
runs contrary to our variability results ($\S$\ref{sec:variability}),
leaving only M2d as our preferred model.

When modeling M2d, we find a best-fit power-law slope of
\hbox{$\Gamma=2.10^{+0.06}_{-0.07}$}, which is marginally higher than
the average AGN value of $\Gamma\sim1.9$
\citep[e.g.,][]{Reeves2000}. Notably, high $\Gamma$ values are often
associated with high Eddington ratio systems
\citep[e.g.,][]{Shemmer2006, Risaliti2009, Brightman2013}, and thus
the slope here is consistent with our initial accretion rate
assessment in $\S$\ref{sec:intro}.  The high-energy cutoff value for
this model, $E_{\rm cut}=128^{+115}_{-44}$\,keV is perhaps somewhat
low. This could imply low coronal temperatures, although the error
bars indicate this value is not well constrained. With this model, we
derive total observed X-ray luminosities of $L_{\rm 2-10\,keV,
  obs}=1.8\times10^{41}$~erg~s$^{-1}$ and $L_{\rm 10-40\,keV,
  obs}=5.6\times10^{41}$~erg~s$^{-1}$, and intrinsic\footnote{This
  does not include contributions from scattered components or
  contamination.} X-ray luminosities of $L_{\rm 2-10\,keV,
  intr}=2.2\times10^{43}$~erg~s$^{-1}$ and $L_{\rm 10-40\,keV,
  intr}=1.5\times10^{43}$~erg~s$^{-1}$, respectively. This intrinsic
$L_{\rm 2-10\,keV}$ value is only a factor of $\approx$1.6 lower than that
predicted by mid-IR to X-ray relation of 
\citet{Gandhi2009}, despite the obvious spectral complexity that
we find.

We stress that the scattered emission from \obj is clearly complex and
thus the models attempted were by no means exhaustive. Alternative
complex component combinations likely exist which can fit the obvious
Compton hump and Fe fluorescence line as well as strike a balance in
the overall reflection continuum levels. Nonetheless, we can conclude
that simple configurations such as a single nuclear
reflector or a patchy torus fail to match the data, and an additional
lower column density component is needed. 

\section{Discussion}\label{sec:discuss}

From the combined modeling we performed in the previous section, there
are a few points worth stressing. 

The quality of the \nustar{} data plays an important role in
constraining the fits. With poorer quality data, such as that from
\suzaku{}, \bepposax{}, or \swift{} BAT shown in
Figure~\ref{fig:M04_fix_allhard}, several of the models we considered
produce acceptable fits.  Only with the \nustar{} data can we observe
in detail the nature of the rising Compton hump and broad peak, which
is difficult to fit with a single cold reflection model. Likewise,
fitting the \chandra{} nuclear and host spectra seperately, we find
that the combination of good-quality nuclear and host spectra creates
considerable tension for several models which would otherwise fit the
total \xmm{} and \nustar{} spectra at acceptable levels. This study
demonstrates that it can be important to have both high-quality spectra
above 10 keV and spatially resolved X-ray spectra in order to, e.g.,
reject simple monolithic cold reflection models. The recent analysis
of the Circinus Galaxy by \citet{Arevalo2014} also benefited from the
powerful combination of high-quality \nustar{} data and spatial
separation of the nuclear and host components, demonstrating that
there too a significant fraction of the warm and cold reflection
components arise from well beyond 2\arcsec{} (i.e., 38\,pc at the
distance of Circinus).

These two objects are among the closest and X-ray brightest
Compton-thick AGN on the sky, and benefit from a wealth of
high-quality X-ray data. Unfortunately, there are only a handful of
nearby Compton-thick AGN where a similar analysis can be made, but it
will be interesting to see how diverse parameter space might be with
respect to this multiple cold reflector model. For fainter and more
distant obscured X-ray AGN, however, we can only obtain modest-to-poor
quality \nustar{} data. Moreover, with the angular resolution of
currently available instruments, we will be unable to separate the
2--8\,keV nuclear emission from its host. So while it may be possible
to model the total emission from such AGN in reasonable detail and
with acceptable results \citep[e.g.,][Brightman et al. 2015 in
  prep]{Balokovic2014, Gandhi2014, Lansbury2014, DelMoro2014}, it will
not be possible to investigate the detailed physical properties of
such sources, as for \obj and Circinus \citep{Arevalo2014}.  The work
here and in Circinus highlight the potential issues of modeling a
total spectrum from, e.g., \xmm{} or \nustar{} with a monolithic model
of the obscurer. For the multiple cold reflector model shown in
Figure~\ref{fig:M2_fullmods}, different portions of the total
reflection spectrum seen by \nustar{} and \xmm{} appear to arise from
different obscuring clouds, decoupling the two key features of cold
reflection. The fact that cold reflectors occur on a variety of
physical scales or with a variety of column densities is unlikely to
change the basic requirement for a high column density associated with
a mildly or heavily Compton-thick AGN. However, it is possible for
this variety to change interpretations regarding the relative Fe
abundance, inclination angle, covering factor for a given column
density, and high-energy cutoff; we observed several of these to vary
significantly from model to model in $\S$\ref{sec:new}.

Although unobscured AGN are dominated by the transmitted power law,
the Fe line and Compton hump do imprint themselves as secondary
contributions. To test how our preferred model of \obj might affect
the fitting of unobscured AGN, we inverted the inclination angles of
the {\tt MYTorus} components by 90\degr{} and added a relativistically
blurred ionized disk reflection component \citep[{\tt
    relconv*xillver};][]{Dauser2013, Garcia2014}.  We linked the disk
reflection parameters to previously determined values (e.g., $\Gamma$,
$E_{\rm cut}$, $\theta_{\rm inc}$, $Z_{\rm Fe}$), or fixed them to
their default values. We normalized the disk reflection relative to
the other components such that it provides the same contribution at 30
keV as the combined cold reflection components. In this configuration,
the relative total reflection flux is high, comprising $\approx30$\%
of the total at 30\,keV, yet the narrow observed Fe K$\alpha$
equivalent width (EW) is only 40\,eV; the latter value is toward the
low end of EW measurements made for Seyfert 1s
\citep[e.g.,][]{Yaqoob2004} and implies that the narrow Fe K$\alpha$
EW may not be a useful estimator for the relative strength of the cold
reflection component, as is sometimes assumed, and even low EW Fe
lines may signify important scattered-light contributions at higher
energies.

We then varied the exponential cutoff energy for our unobscured
version of \obj between three values (100, 300, and 500 keV). We 
simulated a 50\,ks \nustar{} spectrum, resulting in $\sim10^{6}$ 3--79
keV photons, and fit this with a model typical of those used in
unobscured AGN studies (i.e., where the transmitted, disk reflection,
and cold reflection are modeled as ({\tt
cutoffpl+relcov*xillver+pexrav+zgauss},
  respectively, absorbed by a low column density
{\tt tbabs$_{\rm Gal}$}). We allowed
$\Gamma$, $E_{\rm cut}$, $Z_{\rm Fe}$, and the component
normalizations to vary, and fixed the remaining parameters at typical
values (e.g., $Xi=3.1$, $Z_{\rm Fe}=3$, $a=0.9$, $\cos{\theta_{\rm
    pexrav}}=0.3$, $\theta_{\rm xillver}=20$\degr{}). In all cases,
  we obtained reasonable fits with $\chi_{\nu}\approx1.0$--1.1 and
  found that the powerlaw slope was consistent with its input
  value. For input $E_{\rm cut}$ values of 100, 300, and 500 keV, we
  obtained best-fit values of $312^{+43}_{-32}$, $227^{+27}_{-16}$,
  $302^{+37}_{-30}$\,keV, respectively, and $Z_{\rm
    Fe}=(0.7$--$0.8)\pm0.1$. We ran another simulation,
  naively assuming the M2d reflection components were globally the
  same, which yielded similar results for the cutoff energies. Such
  toy models are admittedly far from conclusive due to the likely
  large number of permutations of possible spectral shapes of
  components and degeneracies among parameters, not to mention the
  manner in which we implemented the high-energy cutoff for {\tt
    MYTorus}. Nonetheless, they do highlight how errors on some
  quantities such as the high-energy cutoff could be underestimated
  even in unobscured AGN and can strongly depend on what model
  assumptions are adopted.

The best-fit model for the composite X-ray dataset, M2d, could be
visualized as follows. In the inner 2\arcsec{} (140\,pc) region, we see
a $\theta_{\rm inc}=90$\degr{} (fixed), $N_{\rm
  H}\approx10^{25}$\,cm$^{-2}$ reflector with a covering factor of 0.5
(fixed), which to first order is presumably associated with a
standard, compact, torus-like structure. Additionally, we find a
$\theta_{\rm inc}=0$\degr{} (fixed), $N_{\rm
  H}\approx10^{23}$\,cm$^{-2}$ reflector with an estimated covering
factor of 0.13, based on the relative component normalizations, which
appears to act as a screen. This less dense component could be more or
less co-spatial with the dense torus or it could be material in the
ionization cone. In both cases, we might expect a stratification of
dense material stemming from instabilities associated with the
photoionization of the dense molecular gas by AGN radiation field
structures \citep[e.g., akin to the structures at the boundaries
  between H{\sc ii} regions and molecular clouds;
  e.g.,][]{Pound1998}. Or alternatively, it could simply be reflection
from larger-scale interstellar clouds aggregating within the inner
$\approx100$\,pc \citep[e.g.,][]{Molinari2011}. In all cases, we
should expect a range of clouds which follow a log-normal column
density distribution \citep[e.g.,][]{Lada1999, Goodman2009,
  Lombardi2010, Tremblin2014}. This should in turn introduce
considerable complexity into the AGN reflection components.  We appear
to be seeing the first hints of this anticipated complexity in
\obj. We note that this less dense reflection component produces the
bulk of the Fe K$\alpha$ line emission and, moreover, we see no strong
long-term variability from the $<10$\,keV continuum or line flux. Thus
we conclude that this second reflection component likely arises light
years from the central AGN and/or is distributed enough to wash out
any variability.

We note that at a basic level, the above multi-component reflector
configuration found in the nuclear region appears reasonably
consistent with the picture stemming from mid-IR interferometry for
\obj \citep[e.g.,][]{Jaffe2004, Lopez-Gonzaga2014}, whereby a
three-component model, comprised of a small obscuring torus and two
dusty structures at larger scales (at least 5--10\,pc), best fits the
data. The larger scale dust is off-center and could represent the
inner wall of a dusty cone (e.g., the ionization cone). Based on the
compactness and detailed modeling of spectral energy distributions in
various AGN, these structures are believed to be clumpy and
comprised of a range of torus clouds with column densities of $N_{\rm
  H}\sim10^{22}$--$10^{23}$\,cm$^{-2}$ \citep[e.g.,][]{Elitzer2006,
  Nenkova2008, RamosAlmeida2009}.

On more extended ($>2$\arcsec{}) scales, we find an additional
$\theta_{\rm inc}=0$\degr{} (fixed), $N_{\rm
  H}\approx(4$--$10)\times10^{24}$\,cm$^{-2}$ reflector with a covering
factor of 0.03. The inclination angle, if left free, is not strongly
constrained, and thus it is not clear whether this component is a
screen, a mirror, or perhaps both. This material could be associated
with clumpy molecular clouds either within the ionization cone or the
general interstellar cloud population in the host
galaxy. Intriguingly, our separation of nuclear and host spectra was
purely based on instrumental reasons, and thus, if the distribution of
clouds is strongly centralized and goes roughly as $1/r$ or $1/r^{2}$
\citep[e.g.,][]{Bally1988,Nenkova2008}, then we might expect at least
a fraction of the Fe K$\alpha$ line flux currently assigned to the
$N_{\rm H}\approx10^{25}$\,cm$^{-2}$ torus-like nuclear reflection
component to in fact arise from reflection by extended material.
This suggests that a non-negligible portion of the overall reflection
component in \obj arises outside of the torus. As we found in
$\S$\ref{sec:Fe_line}, the empirical fraction of extended Fe K$\alpha$
flux is substantially higher ($\approx$30\%) than the estimate of the
overall reflection, suggesting that perhaps there are multiple $N_{\rm
  H}$ components responsible for the extended emission as well. Based
on the same molecular cloud distribution argument as above, it may be
possible for the majority of the narrow Fe K$\alpha$ emission to
originate from radii well beyond the classic torus.

\section{Conclusions}\label{sec:conclude}

We have characterized the X-ray spectra of the archetypal
Compton-thick AGN, \obj, using newly acquired \nustar{} data, combined
with archival data from \chandra{}, \xmm{}, and \swift{} BAT. We
modeled \obj with a combination of a heavily obscured transmitted
power law, scattering by both warm and cold reflectors, radiative
recombination continuum and line emission, and off-nuclear point
source emission, employing a handful of cold reflector models. Our
primary results can be summarized as follows:

\begin{itemize}

\item The $>$10\,keV \nustar{} data are consistent with past
  measurements to within cross-calibration uncertainties, but provide
  at least an order of magnitude more sensitivity, allowing us to
  constrain the high-energy spectral shape of \obj in better detail
  than ever before.  We find no strong evidence for short- or
  long-term variability, consistent with the primary transmitted
  continuum being completely obscured from our line-of-sight.

\item We use \chandra{} ACIS-S and HETG data to split the
  reflection-dominated spectrum of \obj into two spatial regimes
  representing the nuclear ($<2$\arcsec{}) and host (2--75\arcsec{})
    contributions to the total spectrum measured by \nustar{}, \xmm{},
    and \swift{} BAT. Because reflection arises from the two distinct
    spatial regimes, modeling both components together allow us to
    break previously unexplored degeneracies to aid physical
    interpretation.

\item Modeling \obj as a monolithic cold reflector with a single column
  density $N_{\rm H}$ generally fails to reproduce some critical
  portion of the combined spectra accurately and/or yields parameters
  which are difficult to reconcile with robust independent
  observations, regardless of the Compton-reflection model used.

\item Modeling \obj using a multi-component reflector (here as
  best-fit model M2d with two nuclear and one extended {\tt MYTorus}
  components with best-fit values of $\Gamma\approx2.1$, $E_{\rm
    cut}\ga90$\,keV, $N_{\rm H}=10^{25}$\,cm$^{-2}$, $N_{\rm
    H}\approx1.5\times10^{23}$\,cm$^{-2}$, and $N_{\rm
    H}\approx5\times10^{24}$\,cm$^{-2}$, respectively, was able to
  reproduce all of the primary spectral lines and continuum shape
  around the Compton hump. In this best-fit multi-component reflector
  model, the higher $N_{\rm H}$ components contributed flux primarily
  to the Compton hump above 10\,keV while the lower $N_{\rm H}$
  nuclear reflector is needed to reproduce the curvature of the
  continuum around 10 keV and it also provides the missing Fe line
  flux to model the whole structure with solar (as opposed to highly
  supersolar) metallicity. Thus, this configuration effectively
  decouples the two key features of Compton reflection which are
  typically assumed to be coupled.

\item There are strong differences in the ratios of the 2--10\,keV
  fluxes of the warm and cold reflection components, depending on the
  model employed and the parameters being fit. Because of the
  decoupling mentioned above, it could be dangerous to
  extrapolate the full properties of the reflector using 
  simple reflection models, as has typically been done in the past with
  either lower-quality data or in type 1 AGNs dilluted by transmitted
  continuum. We note that this decoupling could be at least partially
  responsible for some of the apparently high Fe abundances which have
  been quoted in the literature (e.g., M04).

\item Considering only the \chandra{} data, we find that $\approx$30\%
  of the neutral Fe K$\alpha$ line flux arises from $>$2\arcsec{}
  ($\approx$140~pc) in an extended configuration. Extrapolating this
  fraction inward assuming an increasing solid angle of dense
  molecular clouds implies that a significant fraction (and perhaps
  the majority) of the Fe K$\alpha$ line arises from Compton-scattering
  off of material well outside of the fiducial 1--10\,pc torus
  material. A follow-up investigation looking into the spatial
  distribution of this material around several local AGN will be
  presented in Bauer et al. (2015, in preparation).

\item The multi-component reflector configuration envisioned here
  comprises a compact Compton-thick torus-like structure covering 50\%
  of the sky and more tenuous, extended $N_{\rm
    H}\approx10^{23}$\,cm$^{-2}$ clouds covering $\approx13$\% of the
  sky within the nuclear region ($<$140\,pc), as well as larger-scale,
  low-covering factor Compton-thick clouds which extend out to 100s of
  pc. This scenario bears striking similarities to the multiple dust
  structures found via mid-IR interferometry for \obj, and may
  eventually allow some independent corrobration of the clumpy torus
  model.

\end{itemize}

The benefits of combining high-quality $>$10\,keV spectral sensitivity
from \nustar{} and spatially resolved spectroscopy from \chandra{} are
clear, and could offer novel constraints on the few dozen closest,
brightest AGN on the sky.  Moving on to fainter and more distant
objects, however, is likely to be challenging with current
instrumentation due to the extremely long integrations required and
the increasingly poor intrinsic spatial resolutions obtained.
Moreover, we should caution that our best-fit multi-component
reflector, which we modeled only with three distinct column densities,
could be an oversimplification, and in fact there might be a
continuous distribution of different column-density reflectors, given
that the Galactic molecular cloud probability distribution function is
well represented by a power law over a wide range of column densities
\citep[e.g.,][]{Lada1999, Goodman2009, Lombardi2010}. Each cloud might
contribute something to the overall reflection spectrum, thereby
modifying the spectral shape away from that of a single monolithic
reflector. Hopefully by acquiring similar constraints in other nearby
Compton-thick AGN to those found for \obj and Circinus, combined with
an assessment of the parameter space for obscuring clouds from mid-IR
interferometry studies, we can amass enough clues in the short term to
model distant and/or faint objects in a more informed
manner. Ultimately, if the {\it Athena} mission \citep{Nandra2013,
  Nandra2014} can achieve its best-case scenario for spatial
resolution of a few arcseconds, it could open up spatially resolved Fe
analysis to a significantly larger range of AGN and help us to place
these local AGN in broader context.

\acknowledgements
This work was supported under NASA Contract No.  NNG08FD60C, and made
use of data from the \nustar{} mission, a project led by the California
Institute of Technology, managed by the Jet Propulsion Laboratory, and
funded by the National Aeronautics and Space Administration. We thank
the \nustar{} Operations, Software and Calibration teams for support with
the execution and analysis of these observations. This research has
made use of the \nustar{} Data Analysis Software (NuSTARDAS) jointly
developed by the ASI Science Data Center (ASDC, Italy) and the
California Institute of Technology (USA). This research has made use
of data obtained through the High Energy Astrophysics Science Archive
Research Center (HEASARC) Online Service, provided by the NASA/Goddard
Space Flight Center. We acknowledge financial support from the
following: CONICYT-Chile Basal-CATA PFB-06/2007 (FEB, ET), FONDECYT grants
1141218 (FEB), 1140304 (PA), 1120061 (ET), and Anillo grant ACT1101
(FEB, PA, ET); Project IC120009 ``Millennium Institute of Astrophysics
(MAS)'' funded by the Iniciativa Cient\'{\i}fica Milenio del
Ministerio de Econom\'{\i}a, Fomento y Turismo (FEB); 
Swiss National Science Foundation through the Ambizione
fellowship grant PZ00P2\_154799/1 (MK);
\nustar{} subcontract 44A-1092750 (WNB, BL);
NASA ADP grant NNX10AC99G  (WNB, BL);
ASI/INAF grant I/037/12/0-011/13 (SP, AC, AM and GM);
and STFC grant ST/J003697/1 (PG).

{\it Facilities:} 
\facility{CXO (ACIS, HETG)},
\facility{XMM (pn, MOS)},
\facility{NuSTAR (FPMA, FPMB)},
\facility{Swift (XRT, BAT)},
\facility{BeppoSAX (MECS, PDS)},
\facility{Suzaku (XIS, PIN)},

\bibliographystyle{apj3}

\end{document}